\documentclass[11pt]{article}

\usepackage[margin=1in]{geometry}
\usepackage{graphicx}
\usepackage{amsmath, amssymb}
\usepackage{authblk}
\usepackage{setspace}
\usepackage{hyperref}
\usepackage{booktabs}
\usepackage{float}	
\usepackage{tabularx}
\usepackage{hyperref}
\usepackage[numbers]{natbib}
\usepackage[T1]{fontenc}

\hypersetup{
    colorlinks=true,
    linkcolor=blue,
    citecolor=blue,
    urlcolor=blue
}

\title{Generating a Contact Matrix for Aged Care Settings in Australia: an agent-based model study}

\author[1]{Haley Stone}
\author[1]{Mohana Kunasekaran}
\author[2]{Chris Poulos}
\author[1]{C. Raina MacIntyre}
\author[2]{David Heslop}
\affil[1]{Kirby Institute, University of New South Wales}
\affil[2]{School of Population Health, University of New South Wales}

\date{}
\begin{document}

\maketitle
\onehalfspacing

\begin{abstract}
Understanding infectious disease transmission in institutional settings requires modelling approaches that can represent how contacts arise from structured routines, roles, and spatial constraints. In aged care facilities, interactions are shaped by care delivery processes, staff scheduling, and resident mobility, producing contact patterns that differ fundamentally from those assumed in population-level models. However, standard contact matrices used in epidemiological modelling are typically derived from general population surveys and do not capture these institutional mechanisms.

This study develops an agent-based modelling framework to generate high-resolution contact matrices for aged care facilities by simulating task-driven behaviour, staff workflows, and movement through shared spaces. Rather than prescribing contact structure, interactions emerge endogenously from scheduled activities and proximity during task execution. The model is parameterised using collected activity-diary data from aged care workers and is implemented with behavioural logic decoupled from the physical layout, allowing adaptation to alternative facility designs without modifying core mechanisms.

Simulation results show pronounced heterogeneity in contact patterns across resident care levels and staff shifts. Low and medium care residents exhibited substantially higher contact frequencies than high care residents, while staff working day and afternoon shifts accounted for the majority of resident–staff interactions. Temporal analyses revealed clustering of contacts around structured daily routines, including meals and communal activities. Integrating a proximity-based airborne transmission component parameterised for SARS-CoV-2 demonstrated that transmission risk was concentrated during high-contact shifts and among more mobile resident groups. Vaccination scenarios reduced predicted transmission substantially, with the greatest reductions observed when both staff and residents were vaccinated.

By explicitly linking organisational processes to emergent contact structure, this framework provides a reproducible and transferable approach to contact matrix generation for institutional environments. The model supports more realistic transmission modelling and offers a basis for evaluating targeted infection control strategies in high-risk care settings.
\end{abstract}

\section{Introduction}
Infectious disease transmission within institutional settings presents distinct modelling challenges that are not adequately addressed by approaches developed for general population dynamics \cite{lorig2021abmreview}. In environments such as aged care facilities, interactions are structured by formal roles, scheduled routines, and shared physical spaces, producing contact patterns that differ fundamentally from those assumed in many epidemic models \cite{merler2009role,ferguson2006strategies}. Understanding how these structured interactions give rise to transmission risk requires modelling frameworks that can represent not only who interacts with whom, but how and why those interactions occur.

Aged care facilities are a particularly pertinent example of this challenge. Daily operations are organised around care delivery, staff allocation, and resident needs, resulting in repeated and prolonged interactions between specific individuals \cite{Kimball2020Asymptomatic, rainwater2014influenza, quigley2022covid, mcmichael2020epidemiology, onder2020case, lansbury2017influenza}. These interaction patterns are highly heterogeneous, vary across time and space, and are closely tied to organisational processes rather than individual choice alone. As a result, transmission risk in these settings is shaped by the sequencing of activities, staff workflows, and proximity during task execution, rather than by homogeneous mixing within broad population categories \cite{wallinga2006socialcontacts}.

Many infectious disease models rely on contact matrices to summarise interaction patterns between population groups. While effective for large-scale modelling, such representations typically treat contacts as static, externally specified inputs and abstract away the mechanisms that generate them \cite{mossong2008social}. In institutional environments, this abstraction risks obscuring key drivers of transmission and limiting the interpretability of model outputs \cite{Willem2015Optimizing}. Addressing this limitation requires approaches that can generate contact patterns endogenously from behavioural and organisational processes.

This study develops an agent-based modelling framework to generate contact matrices for aged care facilities in which interactions emerge from routine activities, staff allocation, and movement through shared spaces. Rather than prescribing contact structure in advance, contacts arise from the sequencing of tasks and spatial proximity during care delivery. Behavioural scheduling and task logic are implemented independently of the facility layout, allowing alternative floorplans to be introduced without modifying the underlying interaction mechanisms. The following section situates this approach within existing modelling paradigms and examines common assumptions about contact representation in institutional settings.

\section{Background}

Mathematical and computational models of infectious disease transmission rely critically on how interpersonal contacts are represented. In many modelling frameworks, contacts are summarised using contact matrices that encode the expected number of interactions between predefined population groups over a given time period \cite{mossong2008social}. These matrices are typically treated as fixed inputs, specified exogenously to the model dynamics, and are often derived from self-reported contact surveys conducted over short observation windows. Such representations have proven useful for large-scale population modelling, where fine-grained behavioural detail is neither available nor computationally tractable \cite{wallinga2006socialcontacts}.

However, contact matrices encode strong and often implicit assumptions about how interactions arise. By construction, they abstract away the processes through which contacts are generated, collapsing heterogeneous individual behaviours into aggregate summaries. Temporal ordering, task sequencing, and spatial constraints are typically omitted, and repeated interactions between the same individuals are treated equivalently to independent encounters \cite{Willem2015Optimizing}. These assumptions imply that contact opportunities are weakly structured in time and largely independent of organisational context. While such simplifications may be defensible for modelling diffuse community transmission, they become problematic in institutional environments where interactions are shaped by formal roles, routines, and shared spaces \cite{lorig2021abmreview}.
Aged care facilities exemplify such environments. Contacts within these settings do not arise randomly, nor are they evenly distributed across individuals. Instead, interactions are generated through care routines, staff assignment, task prioritisation, and the physical organisation of space \cite{cuevas2020abmfacilities}. Residents’ contact opportunities depend on care level and mobility, while staff contacts are structured by shift schedules, workload, and responsibility for specific residents or wards. These processes give rise to repeated, temporally clustered interactions between particular individuals, producing exposure patterns that cannot be adequately captured by static or aggregate representations. As a result, models that rely on externally specified contact matrices may mischaracterise both the timing and intensity of transmission opportunities, with implications for estimated outbreak dynamics and intervention effectiveness \cite{Baek2021Correction}.

Empirical contact matrices derived from general population surveys, such as POLYMOD, illustrate these limitations. While widely used, these matrices are based on short recall periods and broad population categories, and are not designed to represent task-driven or institutionally constrained interactions \cite{mossong2008social}. Applying such matrices to aged care settings implicitly assumes that contacts within facilities resemble those of the broader community, differing only in scale rather than structure. This assumption obscures key mechanisms of transmission, including prolonged staff–resident interactions, repeated care encounters, and spatial clustering within shared indoor environments \cite{Kimball2020Asymptomatic}.

Agent-based models (ABMs) offer an alternative approach by representing individuals as autonomous agents whose interactions emerge from behavioural rules, task execution, and movement through space \cite{ferguson2006strategies,eubank2004modelling}. Rather than treating contacts as exogenous inputs, ABMs generate contact patterns endogenously as outcomes of simulated social and organisational processes. This generative perspective allows contact formation to be coupled directly to activity schedules, role-based responsibilities, and spatial constraints, making ABMs particularly well suited to modelling institutional settings \cite{lorig2021abmreview}. Previous studies have used ABMs to generate context-specific contact patterns in schools, workplaces, hospitals, and urban environments, demonstrating that heterogeneity in contact structure can substantially alter predicted transmission dynamics \cite{panovska2022modelling,kerr2021covasim}.
Despite these advantages, ABM-based contact generation approaches often introduce their own limitations. Many models are tightly coupled to specific physical layouts, datasets, or operational assumptions, limiting their transferability and reproducibility \cite{Baek2021Correction}. Behavioural logic is frequently embedded within spatial representations, making it difficult to adapt models to alternative architectural configurations without extensive redevelopment. This coupling constrains the ability to explore how changes in facility design or organisation might influence contact patterns and transmission risk.

From a modelling perspective, this tension highlights a broader trade-off between contextual realism and generalisability. Capturing the mechanisms that generate contacts in institutional environments requires explicit representation of tasks, schedules, and space, yet overly bespoke implementations risk producing models that are informative only for a single setting \cite{lorig2021abmreview}. Addressing this challenge requires modelling frameworks that treat contact formation as a dynamic outcome of behavioural and organisational processes, while maintaining sufficient modularity to support reuse and adaptation across facilities.

The present study adopts this perspective by conceptualising contacts not as predefined inputs, but as emergent properties of task-driven activity within an institutional environment. Behavioural and scheduling processes are implemented independently of the physical layout, allowing alternative floorplans to be incorporated without altering the core logic governing interaction formation. This design choice reflects a commitment to representing the mechanisms that generate contacts, rather than prescribing their aggregate structure, and provides a foundation for reproducible, institution-specific contact matrix generation.

\section{Methods}\label{sec:methods}
In this study, an agent-based simulation is used to estimate contact matrices within a synthetic aged care facility parameterised from empirical data collected in a daily activity survey created for this study. The agent-based simulation model uses the Anylogic 8 software to analyse interactions and contact patterns within an aged care facility, and is designed to simulate the daily activities of residents and staff, capturing the dynamics of movement, task execution, and social interactions. 

The model was developed and documented following the ODD (Overview, Design concepts, Details) protocol for agent-based models \cite{Grimm2010ODD}. Human decision-making processes were additionally specified using the ODD+D extension, which supports the explicit representation of behavioural decision-making in agent-based systems \cite{muller2013oddd}. The model integrates behavioural rules, task prioritisation, and social force algorithms to reflect the operational complexities of a real-world aged care environment.

\subsection{Overview}\label{subsec:overview}
The model represents the aged care facility as a continuous environment, structured into various functional areas such as bedrooms, bathrooms, dining rooms, and communal spaces. The agents in the model are designed with characteristics, features and behaviours relevant to their roles, such as mobility, care level, and shift schedules. The simulation operates over a continuous time scale with repeating schedule every 24 hours, with the model tracking the location, movement, and interactions of all agents throughout the simulation period. As time in simulation progresses, interactions between agents are recorded as contacts, which are defined based on proximity and duration criteria. These contacts are opportunities for disease transmission and can then be used in further disease modelling such as deterministic or stochastic models at larger scale. The model's logic includes the generation, prioritisation, and execution of tasks, which are governed by statecharts, task generation functions, and a pedestrian model that manages agent movement within the environment [Figure 1]. 

\begin{figure}[H]
    \centering
    \includegraphics[width=0.8\linewidth]{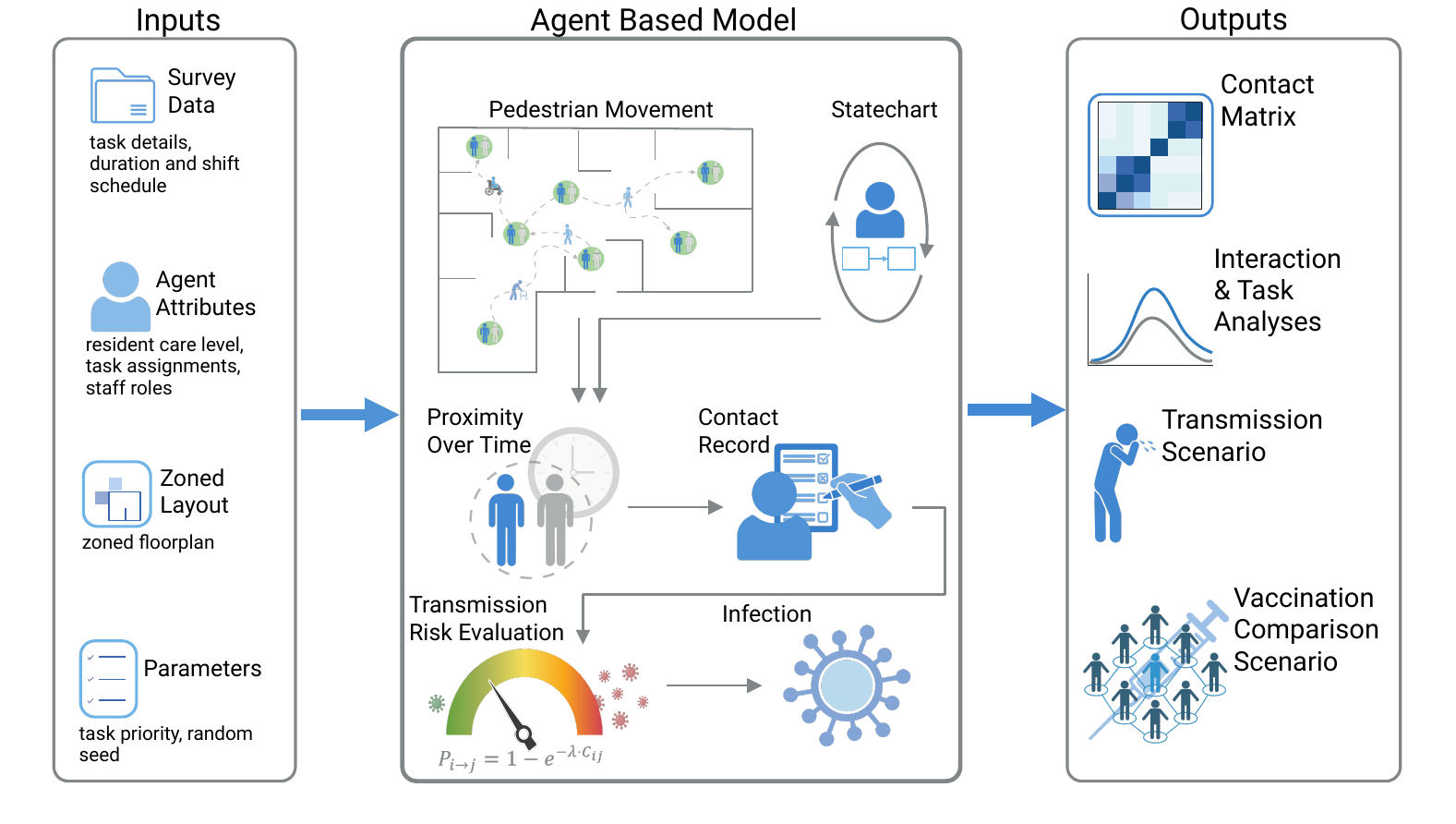}
    \caption{Proposed agent-based simulation pipeline used to model contact patterns and transmission scenarios in Australian aged care facilities. }
    \label{fig1}
\end{figure}

\subsection{Design Concepts}\label{subsec:description}
\subsubsection{Parameterisation}
The behaviours, timings and key parameters governing the day-to-day activities of residents and staff agents in the model was informed through the collection of diarised activity data from aged care workers. This data was also used to develop the suite of behaviours represented within the ABM for the main agent types. The aim of this data collection was to support the development of representative parameters closely mirroring what occurs in real aged care facilities. This data is not available in the peer reviewed literature or publicly available datasets, which are often focused on broader health contacts or interventions. 
The survey comprised of 21 questions of aged care workers informed the model’s parameters, providing insights into regular activities and the frequency and duration of interactions between staff and residents. This information guided the structuring of shift schedules and staff movement patterns within the model. By understanding the frequency and duration of these interactions, the model was parameterised to mirror real-world dynamics. This detailed approach is essential for simulating the nuanced patterns of contact that drive disease transmission in aged care settings. 

\begin{table}[H]
\caption{The estimated durations and their corresponding 95 confidence intervals (CI), timing, and allocation of daily tasks performed by aged care staff, as reported by survey respondents.}\label{tab1}%
\begin{tabular}{@{}p{3.5cm}p{2cm}p{2.0cm}p{2.0cm}p{4.5cm}@{}}
\toprule
\textbf{Task} & \textbf{Duration (min)} & \textbf{Standard Deviation} & \textbf{Respondents (\%)} & \textbf{Details} \\ 
\midrule
Handover & 15 & 0 & 100\% (n=21) & Documentation handover between staff groups. Shift starts 15 minutes before prior shift ends. Times: 6:45, 14:30, 22:15. \\ 
General Observation & 4.67 & 3.03 & 57.1\% (n=12) & No specific time; performed whenever there are no other tasks. \\ 
Aiding in Self Care & 24.17 & 8.78 & 90.5\% (n=19) & Occurs whenever residents need assistance with self-care. \\ 
Meal Prep Breakfast & 5.5 & 1.55 & 100\% (n=12) & Scheduled at 7:45. One staff member from each area takes a break between tasks. \\ 
Meal Preparation & 35.83 & 33.35 & 90.5\% (n=19) & Scheduled at 11:30 and 17:30. Staff take breaks one at a time. \\ 
Mealtime & 38.33 & 22.94 & 90.5\% (n=19) & Times: 8:00, 12:00, 18:00. Over 80\% of residents require aid. Staff assist in dining halls and resident rooms. \\ 
Communal and Personal Activities & 48.33 & 41.68 & 85.7\% (n=18) & Scheduled at 9:00, 13:00, 15:00, and 19:00. Optional, but staff are allocated to assist participating residents. \\ 
Medication Run & 7.67 & 6.34 & 100\% (n=21) & Times: 7:00, 10:00, 14:00, 20:00. Scheduled per residents' needs. \\ 
Breaks & 38.33 & 7.57 & 57.1\% (n=12) & Between hours 6 and 7.5 of a shift. One staff member per area takes a break at a time. \\ 
Cleaning & 12.5 & 9.24 & 81.0\% (n=17) & Scheduled at 9:00, 13:00, and 19:00. Includes post-meal cleaning and periodic 3-hour intervals. \\ 
Laundry & 10 & 9.27 & 74.1\% (n=15) & Scheduled between hours 3 and 4 of shift 1. Staff take turns per area. \\ 
\bottomrule
\end{tabular}
\end{table}

\subsubsection{Contact Definition}
A contact is defined as an interaction in which two agents remain within either 1.5 or 3 metres of each other for duration of at least 3 seconds or more, regardless of whether the contact is physical or non-physical. This duration threshold was used to exclude transient co-presence events that are unlikely to result in meaningful exposure. Contacts are recorded as single continuous events and persist across consecutive simulation time-steps until the agents separate beyond the specified distance threshold.

\subsubsection{Populations}
The model includes two primary categories of agents: i) residents, and ii) staff. The residents are further subcategorised on their mobility and care level needed. The population consists of staff and resident agents, each initialised with attributes relevant to their roles within the simulation. Staff members are assigned to shifts, with their activities structured around the operational needs of the facility. The initialisation process includes assigning residents to specific wings of the facility and organising staff into shifts. These assignments are essential for structuring the flow of activities and interactions within the simulation, ensuring that the model accurately reflects the dynamics of an aged care environment.

The resident agents are categorised into care levels. The three subcategories are defined as (i) high care, which are fully non-mobile to who have very limited mobility and require substantial or event total assistance, such as progressed dementia patients; (ii) medium care, which are residents who have progressed into dementia but not fully immobile and have a moderate degree of independence requiring some assistance; and (iii) low care, which are residents who are very early in their onset and are fully mobile and broadly independent.

In the model, staff agents play a vital role in maintaining the daily operations of the aged care facility, providing care and support to residents across all care levels, high care, medium care, and low care. Unlike residents, staff agents are not divided into categories, but instead have a responsible wing, which has varying resident care levels. Staff agents are scheduled to work in shifts, ensuring continuous care throughout the day and night. The simulation models their movement within the facility, interactions with residents, and execution of tasks, reflecting the real-world dynamics of caregiving in an aged care environment. This approach ensures that the needs of residents at all care levels are met efficiently and effectively, contributing to the overall well-being of the facility’s population.

\subsection{Details}
\subsubsection{Simulation of Physical Environment}
The simulation environment represents a single-storey aged care facility, parameterised from a representative real-world layout to support realistic movement and interaction patterns. The facility's layout is parameterised over an area of approximately 500x500 meters, with specific nodes representing essential locations such as resident bedrooms, bathrooms, dining areas, communal spaces, and staff rooms. These nodes serve as reference points within the simulation, facilitating the movement and interaction of agents throughout the facility.

The spatial layout is implemented as a standalone model component, allowing alternative floorplans to be incorporated by modifying only the layout structure, while agent behaviour, scheduling logic, and contact-generation processes remain unchanged.

\subsubsection{Initialisation}
The simulation is initialised by configuring the environment to more accurately represent the physical layout of the aged care facility, with functional spaces such as bedrooms, bathrooms, dining rooms, and activity areas. Residents and staff are assigned specific locations based on their care levels and shift schedules, aligning with the real-world operations of the facility. Key operational parameters, including resident numbers and shift timings, are initialised to reflect the facility’s daily dynamics as outlined in Table \ref{tab2}, establishing the foundational conditions under which the model operates.
During this initiation phase, decision rules and algorithms are activated to define agent behaviours, such as task prioritisation and interaction patterns. These rules, informed by the survey data and real-world parameters, guide the agents’ decisions throughout the simulation, ensuring that their actions align as closely as possible with real world operational dynamics of the aged care facility. During initialisation, behavioural rules and task priorities are activated to govern agent actions throughout the simulation.

The model incorporates behavioural programming, time-based variability, and parameter variation to simulate the real-world diversity in agent behaviours and outcomes. Key operational parameters, such as staffing levels, shift durations, number of staff per wing, and initial resident population, were systematically varied to explore their influence on contact patterns and on the system.  To reflect the unpredictability of human behaviour, stochastic elements introduce randomness into decision-making processes. A pseudorandom number generator is employed with a new seed number for each modelling run, ensuring that each run is non-replicable and introduces unique variations into the simulation. This randomness, combined with the structured variability, allows the model to capture a wide range of probabilistic interactions and outcomes.

Population distributions and parameters were defined based on empirical data drawn from the survey of staff members, reference to the academic literature, or reasonable assumption following Subject Matter Expert consultation within the aged care industry or specialist clinicians. These settings ensure that the simulation captures a range of probabilistic interactions and outcomes and allow for emergent dynamics to be observed.

To reflect real-world variability in agent behaviour, delay times, and movement, the simulation was run stochastically across 100 replicates per contact scenario. Random seed values were not fixed during these replications to preserve natural variability. The structure and purpose of this replication framework are detailed in the ‘Stochastic Simulation Framework’ section \ref{subsec:stochsim}. However, for intervention transmission simulations, matched random seeds were used across corresponding runs to isolate intervention effects.

\begin{table}[H]
\caption{Model-level parameters fixed at initialisation and applied globally throughout the simulation.}\label{tab2}%
\begin{tabular}{@{}p{5cm}p{2.5cm}p{7.5cm}@{}}
\toprule
\textbf{Parameter} & \textbf{Value} & \textbf{Rationale and Description} \\ 
\midrule
Simulation Start Time & 6:15 & Chosen to allow time for the model initiation to occur before the start of Shift 1. \\ 
Contact Policy Range & 1.5 \& 3 metres & The range within which contact between individuals is considered in the simulation. A contact can be both physical or nonphysical. \\ 
Shift 1 Start Time & 6:45 & Aligns with the morning routine from the staff survey performed in this model. \\ 
Shift 2 Start Time & 14:30 & Aligns with the afternoon routine from the staff survey performed in this model. \\ 
Shift 3 Start Time & 22:25 & Aligns with the night routine from the staff survey performed in this model. \\ 
Number of Staff Shift 1 & 12 & Allows for 2 staff members to be present in each resident wing. \\ 
Number of Staff Shift 2 & 12 & Allows for 2 staff members to be present in each resident wing. \\ 
Number of Staff Shift 3 & 6 & Allows for 1 staff member to be present in each resident wing. \\ 
Number of Residents & 60 & Fills rooms within the physical layout of the facility. \\ 
\bottomrule
\end{tabular}
\end{table}

To evaluate the robustness of emergent contact patterns under varying operational assumptions, the model was reparameterised across a constrained set of plausible configurations. Three input parameters were varied independently: (i) simulation start time (6:00, 6:15, 6:30), (ii) resident population size (45 or 60), and (iii) staff numbers per shift, reflecting 1, 2, or 3 staff per ward across six wards. Each configuration was subjected to 100 independent stochastic iterations, with all other structural and behavioural parameters held constant. Output metrics included total contact count, mean contact duration, and summary properties of the resulting contact networks, enabling assessment of the sensitivity of contact formation dynamics to alternative operational scenarios.

\subsubsection{Agent Activities and Scheduling}
The activity and schedule management within the simulation is structured around a framework that dynamically governs both staff and resident activities, ensuring that all tasks are systematically generated, prioritised, executed, and completed. The system relies on a combination of statecharts, time-based task generation functions, activity tokens, process models to manage task generation and execution, incorporated into a pedestrian model to simulate physical interactions. These components work together to allow for complex interactions to emerge by implementing certain degrees of individual autonomy, accounting for relevant interactions and the environment within the simulation, ensuring that each agent's actions align with their roles and the overall objectives of the system. The activity workflow combines a process model, which initiates and schedules tasks, with a statechart that governs the agent’s active task execution [Figure 2].

Activities within the simulation are initiated and managed through statecharts, which are tailored to each agent type. Each agent is equipped with multiple statecharts that operate simultaneously, each containing conditional information regarding the timing and duration of various behaviours and tasks. These statecharts regulate the flow of activities by determining when an agent should start a task, continue with an ongoing task, or transition to a different state based on evolving conditions within the simulation. Transitions within these statecharts are activated by specific conditions or events, such as the availability of a new task, the completion of a current task, or external triggers. These triggers, or signals, may be time-based, message-based, or condition-based, and the model continuously monitors for them to decide when a new activity should begin. When a trigger condition is met, the statechart transitions the agent from an idle or waiting state to an active state and triggers the generation of the activity function.

To prevent indefinite queuing of low-priority tasks under sustained workload, the scheduling process includes a queue timeout rule. Tasks that remain queued for more than 40 minutes without execution are either escalated in priority or terminated depending on task type, ensuring bounded waiting times and stable system behaviour.

\subsubsection{Activity Generation and Execution}
Activity generation for both staff and resident agents is managed through specific functions invoked by the statecharts or other elements of the simulation logic. The activity generation process involves defining both the activation of the specific activity, including the prioritisation and the creation of its token. Tokens appear only in the components of the process model that actively manage an activity’s lifecycle, such as generation, queuing, prioritisation, execution, and therefore do not appear in statechart states or pedestrian movement blocks. Within the agent a process model reads the generation function and locates the activity’s priority. Both staff and resident models utilise priority-based functions to determine the order of task execution. In addition, activity tokens are employed within the model to represent the discrete units of work or actions that an agent must perform. Each token encapsulates the details of a specific activity, including the required subordinate actions, the sequence in which they should occur, location of activity and any associated timing constraints. Tokens serve as structured carriers of activity-specific information, ensuring that agents execute tasks according to predefined parameters in a consistent manner across the population. The management of activity tokens involves queuing, prioritisation, and execution, with tokens queued based on their priority. Each agent’s process model selects the appropriate token for execution, guiding the agent’s actions and ensuring consistency and adherence to the activities’ requirements. 

While the process models in staff and resident activities integrate statecharts, activity generation functions, and activity tokens to determine how activities are prioritised, queued, and executed based on the agent’s current state and the overarching simulation logic, the process models within the agent’s activity classes, such as Resident Activity or Staff Activity, serve as the frameworks that govern how activities are managed and executed by these agents within the simulation. These processes oversee the lifecycle of activities, from their generation to their completion, ensuring a structured approach to activity management. The interaction between process model within the agent’s activity class, such as Resident Activity or Staff Activity agent, and predefined agent activity templates is another key aspect of activity management. These templates serve as blueprints for specific activities, detailing the steps, resources, and conditions required for their execution within the pedestrian model. When an activity is generated and assigned to an agent, the process model references the corresponding activity template. This template provides a detailed roadmap for the task, including the sequence of actions, timing, and any conditions that must be met for successful execution. Templates are triggered by the process model based on the specific type of activity and the agent’s current state, ensuring that activities are executed consistently and according to predefined standards.

\subsubsection{Social Force Model and Pedestrian Dynamics}
A pedestrian simulation using the well characterised Helbing-Molnár-Farkas-Vicsek social force models is used to implement the movement of agents within the simulation environment \cite{Helbing2000Panic}. This model differs from other social force models by incorporating a more comprehensive set of interaction forces, including individual desires to reach specific destinations while avoiding collisions with other pedestrians and obstacles. It also accounts for the variability in pedestrian behaviour, making it particularly effective for simulating realistic movement patterns in complex environments \cite{Helbing2000Panic}. The pedestrian model within the simulation is triggered when an activity template requires an agent to navigate to a specific location, such as moving between rooms or transitioning to different areas. The process model within the staff or resident activity generates movement details as part of the task parameters, including the destination, path, and required speed. These parameters are passed to the pedestrian model through the activity templates, which handles the agent’s physical navigation. The pedestrian model ensures realistic movement by managing timing, sequence, and interactions with obstacles and other agents. Using a social force approach, it calculates desired speed, directs movement toward attraction points, and programs agents to follow the shortest routes while avoiding collisions. These features collectively enhance the accuracy of agent interactions and behaviours within the simulation.

\subsubsection{Subcomponent Integration}
Once an activity token is selected for execution, the process model coordinates task progression by referencing the associated activity template and invoking the pedestrian model where movement is required. Agent state is updated throughout execution to reflect task progress, interruptions, or completion, ensuring consistency between spatial movement and task logic.

\subsubsection{Feedback Mechanisms}
The simulation also incorporates feedback mechanisms that monitor the execution of tasks and the overall performance of agents. Feedback is collected on key metrics, such as task completion times, movement efficiency, and resource usage. The process model utilises this feedback to adapt future task generation and execution, ensuring that the simulation remains responsive to changing conditions. These adaptive mechanisms might include adjusting task priorities, modifying movement paths, or reallocating resources to optimise performance.

\begin{figure}[H]
    \centering
    \includegraphics[width=1\linewidth]{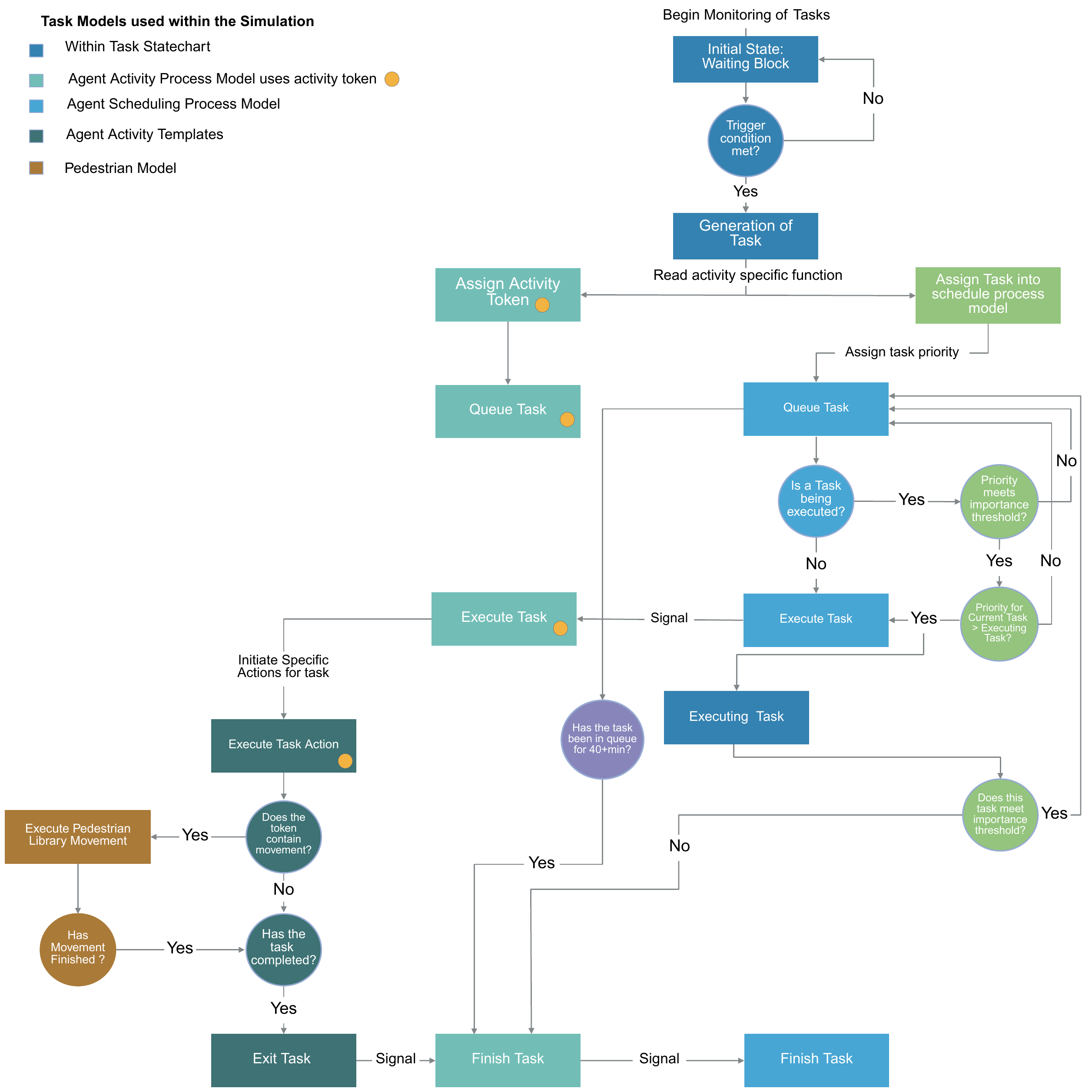}
    \caption{Workflow for task generation, prioritisation, and execution within the agent-based model. Tokens represent individual activity instances as they move through the scheduling process. Blue blocks denote scheduling and control logic, while brown blocks represent physical execution handled by the pedestrian model. Arrows indicate the direction of task flow.}
    \label{fig2}
\end{figure}

\subsubsection{Integrated Model Operation}
While the general framework for activity management applies to both staff and resident agents, there are distinct nuances and additional complexities involved in both resident and staff agent activities. Staff activities within the simulation are designed to reflect the operational responsibilities and task management inherent to the roles of staff agents. 
In addition to the priority-based system, staff agents are assigned to specific residents within their allocated wing or area of responsibility. Each staff agent is designated to manage activities and provide care for residents in a particular wing, ensuring a structured and organised approach to task allocation. The allocation of staff to specific wings allows for optimal task distribution, ensuring that each staff member can focus on the residents within their designated area without the inefficiencies that might arise from constantly shifting between different locations or managing varying levels of care across different resident groups.

For staff, the process models are tailored to manage multiple, often concurrent, activities that may have conflicting priorities. Staff agents are frequently tasked with balancing operational duties, such as logistical support, administrative functions, and direct care activities, in addition to balancing unscheduled events [Table \ref{tab3}]. The process model is designed to adjust execution of tasks as they emerge through priority-based queues, ensuring that tasks are addressed in a logical order based on urgency and importance. Each task is assigned a priority level, and the process model dynamically manages the priority queue, continuously re-evaluating tasks as new ones are generated, or existing tasks are completed. This process model also allows the model to effectively handle conflicting tasks, ensuring the lower priority tasks are either delayed or requeued as necessary. This is particularly important as staff members often juggle multiple operational responsibilities that may require immediate attention. In addition, an importance threshold was assigned in the model to manage the prioritisation-based system. Tasks can be interrupted if they exceed the threshold and are higher priority than the current executing task [Table \ref{tab3}]. If the interrupted task remains within the importance threshold, it is requeued for later completion. The same priority system also contains an Emergency Response category for both staff and residents, which covers rare acute events such as clinical incidents or facility-wide alerts and would supersede all other tasks if triggered. No emergency triggers were scheduled in the scenarios analysed here, so Emergency Response tasks did not occur.

\begin{table}[H]
\caption{Staff Tasks Ranked by Priority Assignment in the Model}\label{tab3}%
\begin{tabular}{@{}p{3.5cm}p{4.9cm}p{1.5cm}p{5.0cm}@{}}
\toprule
\textbf{Task} & \textbf{Description} & \textbf{Priority} & \textbf{Rationale for Priority} \\ 
\midrule
Emergency Response & Immediate response to emergencies such as health crises or accidents involving residents or within the setting. & 100\textsuperscript{†} & These tasks are critical to the safety of the aged care home and residents and must be addressed immediately. \\ 
Medication Administration & Performing administration of medication at allotted times. & 90\textsuperscript{†} & Residents are dependent on punctual medication, and delays could result in health risks. \\ 
Assisting Daily Resident Activities & Aiding in tasks such as eating, dressing, or hygiene-related tasks. & 60--80\textsuperscript{†} & Residents are often dependent on staff to perform these tasks, and delays could impact daily function. \\ 
Self-Hygiene & Staff taking the time for personal hygiene, such as using the bathroom. & 55\textsuperscript{†} & Necessary for staff but still postponed over more pressing tasks listed above. \\ 
Shift Changes and Administration & Managing shift changes, including the handover of responsibilities and updating records. & 50\textsuperscript{†} & Ensures smooth transitions between shifts and smooth simulation operation. \\ 
Group Activity Coordination & Aiding residents during group or communal activities. Staff are assigned tasks based on responsibilities and specific activities. & 40 & Important for resident well-being but less critical than direct care tasks. \\ 
Breaks and Downtime & Breaks are staggered to ensure more than one staff member is present within a working area at all times. & 30 & Required for all staff but scheduled around critical tasks. \\ 
Cleaning Tasks & Perform routine cleaning of resident areas and laundry. & 20 & Necessary for maintaining a clean environment but can be deferred if more urgent tasks arise. \\ 
General Observation and Routine Resident Checks & Regular checks to monitor residents’ well-being. Performed when other tasks are not scheduled. & 10 & Important for early detection of issues but can be scheduled flexibly around higher-priority tasks. \\ 
\bottomrule
\end{tabular}
\footnotesize \textsuperscript{†} Tasks exceeding the importance threshold are requeued into the process model.

\end{table}

For residents, activities are triggered and managed according to their care level, with specific processes determining how and when tasks are initiated. The model incorporates delay timers to introduce variability in activity start times, modulating these delays based on care level to more accurately reflect the balance between required structure and necessary flexibility. For example, residents with medium care needs may experience longer or more frequent delays before starting an activity, simulating the additional time required for assistance or preparation. Additionally, the model accounts for the possibility of residents refusing or delaying certain tasks, particularly in high care scenarios where limited mobility might lead to hesitation or non-compliance. This refusal is integrated into the decision-making process, ensuring that the simulation can adapt to such occurrences.
Care level significantly influences not only the timing of activities but also the type and scope of activities available to residents [Table \ref{tab4}]. Residents with lower care needs have access to a broader range of activities, often with minimal supervision, allowing for more autonomy in their daily routines. In contrast, residents with higher care needs are restricted to activities that are either essential for their well-being or are closely supervised due to their physical and cognitive limitations.

Social interactions are also integrated into resident activities through specific functions that manage the involvement of friends, ensuring that social dynamics are incorporated into both planning and execution. These interactions are particularly important in medium and low care scenarios, where residents are encouraged to engage in social activities to maintain or enhance their well-being. The model facilitates these interactions by assigning preferred locations for activities, which are based on both care level and individual preferences. For example, low-care residents might choose from a variety of social settings, while high-care residents are guided to more accessible or supervised areas.

The pedestrian model manages residents' navigation and interactions within the environment, ensuring that movement between activities is smooth and reflects the residents' abilities. The execution phase of activities considers timing and duration, which are adjusted according to care level and task requirements. Upon completion, residents are transitioned to a neutral state or prepared for subsequent tasks. This transition is carefully monitored to ensure that residents are ready for their next activity, with feedback mechanisms in place to adjust the simulation based on observed resident behaviour and changing environmental conditions.

\begin{table}[H]
\caption{Resident Tasks and Differentiation in Execution Between Care Levels}\label{tab4}
\begin{tabularx}{\textwidth}{@{}p{2cm}p{4.5cm}X X X@{}}
\toprule
\textbf{Task} & \textbf{Function in Model} & \multicolumn{3}{c}{\textbf{Differentiation Between Care Levels}} \\
\cmidrule{3-5}
& & \textbf{Low Care} & \textbf{Medium Care} & \textbf{High Care} \\ 
\midrule
Sitting & If during shift 1 and 2, this is the task if there is nothing else scheduled. The resident will pick from a selection of locations within the model where they prefer to sit. & Residents choose freely from a broad range of locations. & Residents have a more limited set of comfortable or easily accessible locations. & Residents are restricted to locations closer to their assigned wing or areas with higher staff presence. \\ 

Eating & Residents choose to eat in the dining room or other areas depending on the meal and their care level. The model directs them to their preferred or most appropriate eating location based on these factors. & Residents typically eat in the dining room with more freedom to choose alternate spots. & Residents primarily eat in designated dining areas with some flexibility. & Residents are restricted to specific dining locations that are close and supervised. \\ 

Self-Hygiene & Residents getting ready in the morning often take a while and require staff assistance. & Minimal assistance needed; residents are mostly independent. & Some assistance required, with scheduled help from staff. & Significant assistance needed, with frequent staff involvement. \\ 

Communal or Personal Activity & Residents have multiple optional scheduled activities throughout the day, which may include group activities or personal hobbies. The model assigns activities based on care level and personal preferences. & Active and has all activities available. & Fewer options and moderate participation, with some activities adapted to their needs. & Limited participation, mostly in supervised or close-proximity activities or within living space. \\ 

Nap & The model schedules naps based on individual resident preferences and their care levels, often in the afternoon. Residents move to their assigned napping locations, usually their rooms. & Flexible nap times and locations, with more freedom to choose. & Scheduled nap times with more predictable routines. & More strict scheduled naps in specific locations, usually with staff assistance. \\ 

Sleep & Residents have a general schedule of when to go to their room and sleep. & Flexible bedtimes with minimal checkups. & More structured sleep times with occasional staff checks. & More structured sleep times with more frequent staff checks. \\ 

Emergency Response & Residents are to stop moving in the model until staff have deemed the emergency is over. Residents also look for messages from staff to move if the emergency requires an exit from a location. & \multicolumn{3}{c}{No difference between care levels.} \\ 
\bottomrule
\end{tabularx}
\footnotetext{Note: The differentiation in care levels highlights the varying levels of staff involvement and resident autonomy across tasks.}
\end{table}

\subsection{Stochastic Simulation Framework}\label{subsec:stochsim}
To account for inherent randomness in agent behaviour, movement paths, and contact formation, the simulation was replicated 100 times under each scenario. Each run used a consistent task structure, population distribution, and facility layout, but introduced stochastic variation through delay timers, probabilistic task triggers, and agent movement dynamics, as described in the above sections. Random seeds were not fixed across replicates in baseline contact scenarios, allowing emergent variation in contact formation and interaction patterns across runs. However, for transmission and transmission intervention simulations, identical random seeds were used across corresponding simulation iterations, ensuring that agent schedules, behaviours, and movement paths remained consistent between intervention conditions, enabling direct attribution of outcome differences to the intervention effect rather than stochastic variability.

\subsection{Contact Matrix Generation}\label{subsec:contactmatrix}
\subsubsection{Identifying Simulated Contacts}
We simulated the movement of all the agent types, capturing their interactions and proximity throughout each staff shift. This enabled us to directly compute the instances of close contact among all individuals present, along with the duration of these interactions. In addition, as multiple studies on respiratory illnesses revealed that droplets extend far beyond 1.5 metres which is the recommended social distancing policy for the World Health Organization \cite{bourouiba2014sneeze,bahl2022precautions,xie2007droplets}. As a result, separate contact analyses were performed for contact radii of 1.5 and 3 metres to analyse differences in nonphysical contacts within the space. 
In the model, a contact is defined as an event where two agents come within 1.5 or 3 metres of each other for a duration equal to or exceeding 3 seconds. Mathematically, a contact between two agents i and j is represented as:
\[
C_{i,j} = \left[ \text{dis}_{i,j}, \text{dur}_{i,j} \right]
\]
\begin{center}
where $i$ and $j$ are two agents; $\text{dis}_{i,j}$ is defined by the effective radius ($r_e$) of each individual agent, which we defined as 1.5 or 3 metres, has active agents within it; and $\text{dur}_{i,j}$ is a time-step ($t$) by which the two persons interlinked exceeds $(t+3)$.
\end{center}

Within the model, a contact is recorded only when two agents are co-located within the same room and within the specified proximity threshold, preventing spurious contacts across physical barriers. Contacts persist across consecutive simulation time steps until the agents separate, after which any subsequent re-encounter is logged as a new contact event. If the same agents come into contact again after a period of separation, a new contact record is created and logged for that pair of agents. This new contact record will have a new start timestamp. The process continues in this manner, with new contact records being generated for each instance of contact between the same pair of agents, and timestamps being updated accordingly. The contact record, once created, includes the following parameters: (i) the identifier of the contacter; (ii) the identifier of the contactee; (iii) the start time of the contact within the effective radius; (iv) the end time of the contact; and (v) the duration of the contact. 

Once the contact records are generated by the simulation, the total number of contacts for each individual agent is then calculated for each shift. These totals are first summarised into a pairwise individual-level contact matrix of size mxm,  created by aggregating all interactions between each individual agent and every other agent in the facility. This is defined as: 
\[
M_{i,j} = \sum_{j} C_{i,j} \quad \text{for all } j
\]
\begin{center}
where $M_{i,j}$ represents the total contacts between individual $i$ and $j$.
\end{center}

These individual totals are then used to compute an individual-to-type contact matrix of size mxn, which captures the average number of contacts that each individual agent i has with agent type k and is defined by: 
\[
A_{i,k} = \frac{1}{N_k} \sum_{j \in \mathcal{G}_k} M_{i,j}
\]

\begin{center}
where $i$ denotes a focal individual agent; $k$ denotes an agent type; $\mathcal{G}_k$ is the set of agents belonging to type $k$; $N_k = |\mathcal{G}_k|$ is the number of agents of type $k$; and $M_{i,j}$ represents the total number of contacts between individuals $i$ and $j$.
\end{center}

Finally, the individual-level contact values are averaged across agents of the same type to construct the agent type–level contact matrix of size nxn, representing mean contact frequencies between agent types. To derive the nxn matrix, a Generalised Linear Model (GLM) as fitted using a Poisson distribution with a log-link function to model the contact count data. The response variable was the number of observed contacts between each contactee, contacter pair, aggregated by their assigned agent categories. Both predictor variables, contactee category and contacter category, were modelled as categorical fixed effects. The GLM captured the average contact frequencies between agent types, adjusting for variation in group sizes and individual-level heterogeneity in contact formation. Model fitting was conducted separately for each contact simulation replicate. The predicted counts from each fitted model were then extracted and averaged across the 100 stochastic replicates to obtain a stable, smoothed estimate of contact intensity. Corresponding 95\% confidence intervals were computed from the distribution of predicted values across replicates, allowing for uncertainty to be incorporated in the final contact matrix. To capture uncertainty in contact frequency estimates, the GLM-predicted contact matrices from all replicates were combined, and 95\% uncertainty intervals were calculated using the 2.5th and 97.5th percentiles across runs.

\subsubsection{Additional Analyses}
Additional analyses were conducted on staff and resident contacts using timeseries analysis, duration plots, and task analyses to comprehensively assess interaction patterns and task execution within the facility. For the timeseries analysis, contact data was categorised based on predefined agent types and filtered using rolling summations with a 20-minute window to smooth short-term fluctuations and highlight longer-term trends. A separate analysis focused on the duration of contacts within these 20-minute windows. Additionally, an activity time plot was created to examine the tasks completed by residents, including the locations where these tasks occurred within the model.

\subsection{Transmission Risk Simulation}\label{subsec:transmission}
 To complement the construction of contact matrices and provide a framework for interpreting their epidemiological implications, a probabilistic airborne transmission model was integrated into the simulation to calculate infection risk based on agent proximity and cumulative contact duration. Transmission probability was dynamically evaluated for each susceptible-infectious agent pair during the simulation based on recorded spatial proximity and accumulated exposure time. Infection outcomes were determined based on the calculated transmission probability at the end of each recorded contact event. Within this model, the transmission component is parameterised for SARS-CoV-2, employing airborne quanta-based transmission estimates and vaccine effectiveness values specific to this pathogen.

\subsubsection{Initial Seeding}
To initiate the transmission simulations within the model, an initial infection seeding strategy was employed. For each simulation run, a single staff agent was randomly selected at the beginning of a designated shift to serve as the primary infection case. Only one initial infection was seeded per 24-hour simulation cycle. Separate simulation sets were conducted independently for each shift. Each index case was assumed to be infectious from the start of their respective shift, and all infection events were recorded across the 24-hour simulation period. To allow for direct comparison between scenarios, identical random seeds were used across intervention runs, ensuring that all shared stochastic processes, such as agent movement, task allocation, and seeding, remained consistent, isolating the effect of the intervention. Each simulation run included one infectious staff member as the index case for the 24-hour period.

\subsubsection{Transmission Dynamics}
The transmission dynamics presented here are parameterised for SARS-CoV-2, using quanta-generation and inhalation parameters specific to this pathogen. The same formulation can be applied to other airborne diseases by substituting the corresponding pathogen-specific parameters. A probabilistic transmission risk model, applying a proximity-based Wells-Riley formulation, was integrated into the simulation \cite{riley1978measles}. Transmission risk between agents was calculated based on direct proximity and cumulative contact duration, without assuming homogeneous aerosol mixing or dilution effects through room-scale ventilation. The transmission probability $P_{i,j}$ from an infectious individual i to a susceptible individual j was defined as: 
\[
P_{\text{transmission}}(i,j) = 1 - e^{-\lambda t_{i,j}}
\]
\begin{center}
Where $\lambda$ is the effective transmission rate per second of close contact, calculated as the product of the quanta generation rate (\textit{I}) and the inhalation rate (\textit{q}) and normalised to a per-second basis; $t_{i,j}$ represents the cumulative contact duration, measured in seconds, between individuals \textit{i} and \textit{j}, derived directly from simulation-recorded interaction data [Table \ref{tab5}]    
\end{center}

\begin{table}[H]
    \centering
    \caption{Parameter values used in the proximity-based Wells-Riley model}\label{tab5}
    \begin{tabularx}{\textwidth}{lXlX}
        \textbf{Parameter} & \textbf{Description} & \textbf{Assumed Value} & \textbf{Source} \\
        \hline
        \textit{I} & Quanta generation rate & 25 quanta/hr (95\% CI: 14–54) & \cite{buonanno2020quantitative,morawska2020airborne,dai2020ventilation} \\
        \textit{q} & Inhalation rate of susceptible individuals during sedentary activities & 0.5 m\textsuperscript{3}/hr (95\% CI: 0.48–0.54) & \cite{buonanno2020quantitative,guo2020aerosol,usepa2011handbook} \\
        $t_{ij}$ & Cumulative interaction duration between infectious individual $i$ and susceptible individual $j$ & Simulation-derived & – \\
        $\lambda$ & Effective transmission rate per hour of close contact & Computed from $I \cdot q$ & – \\
    \end{tabularx}
    \label{tab5}
\end{table}
The inhalation rate (\textit{q}) was assigned based on discrete proximity thresholds, consistent with the contact definitions within the simulation framework. For interactions recorded at a distance of 1.5 metres, a full inhalation rate of 0.5 m\textsuperscript{3}/hour was applied. Interactions occurring beyond the radius were not considered capable of resulting in transmission and were excluded from transmission probability calculations [Table \ref{tab5}].

\subsubsection{Integration into the Simulation Framework}
The transmission risk formulation described previously was operationalised within the simulation environment by dynamically evaluating infection probabilities at the level of individual agent interactions during each simulation run. Transmission probability calculations were implemented as an additional process executed alongside the existing contact tracking mechanisms.
At each simulation time-step, the model updated the cumulative contact duration t\textsubscript{i,j}  for each susceptible-infectious agent pair that satisfied the defined spatial proximity criteria. Transmission probability was not evaluated independently at each second but was instead determined as a cumulative risk based on the total exposure time recorded over the duration of contact. At the end of each contact event, the cumulative probability of infection was computed, and a single uniform random number $U \sim \mathcal{U}(0,1)$ was drawn for each susceptible agent with a recorded exposure. If the random draw was less than or equal to the computed transmission probability, the susceptible agent was reclassified as infected. Newly infected agents remained infectious for the remainder of the 24-hour simulation cycle, with no secondary transmission events modelled within the same simulation period. Infection status for each agent and cumulative infection counts were recorded at the end of each 24-hour simulation period for subsequent analysis.

\subsubsection{Vaccination Intervention Scenarios}
Given the elevated vulnerability of aged care settings and the recognised role of booster vaccination in mitigating infection risk, intervention scenarios incorporating resident and staff booster coverage were evaluated. Vaccination status was incorporated into the simulation as an agent attribute assigned at initialisation. For each simulation run, staff agents were assigned a binary vaccination status according to the intervention scenario being evaluated. Vaccination status remained fixed throughout the 24-hour simulation cycle. In this study, the baseline condition (“no booster administered”) represents a population whose immunity from the primary vaccination series has waned and is therefore modelled as having no effective protection at the time of exposure.

During the evaluation of transmission probabilities, vaccination status was incorporated as a scaling adjustment applied to the calculated probability of infection for susceptible agents. If a susceptible agent was recorded as vaccinated, their computed transmission probability $P_{\text{transmission}}(i,j)$ was adjusted as:
\[
P_{\text{transmission}}^*(i,j) = P_{\text{transmission}}(i,j) \cdot (1 - \text{VE})
\]
\begin{center}
    where VE is the vaccine effectiveness.
\end{center}
Booster vaccination was modelled assuming receipt of the Pfizer-BioNTech Comirnaty Omicron XBB.1.5 formulation. This vaccine was selected as it represented the predominant COVID-19 booster administered in Australia during the study period, consistent with Australian Technical Advisory Group on Immunisation recommendations prioritising mRNA-based vaccines for booster doses and confirmed in national policy documents and vaccine rollout reports \cite{atagi2025booster,ausgov2025guidance,tga2025approval}. Vaccine effectiveness against any SARS-CoV-2 infection for the Comirnaty Omicron XBB.1.5 booster was assumed to be 70.6\% (95\% CI: 43.3\%–84.8\%), based on recently reported effectiveness estimates encompassing both symptomatic and asymptomatic infections \cite{ward2025comirnaty,tartof2024estimated,caffrey2024effectiveness}. Vaccination status modified the probability of infection at the time of contact evaluation without altering other agent attributes, including mobility patterns, task execution, or contact behaviours.

Within the simulation, vaccination status modified transmission probability only at the point of contact event evaluation. At the conclusion of each cumulative contact between an infectious and susceptible agent, the adjusted transmission probability $P_{\text{transmission}}^*(i,j)$ was computed based on the susceptible agent’s vaccination status. A uniform random number $U \sim \mathcal{U}(0,1)$  was then drawn to determine infection status, following the method described above. Three vaccination intervention scenarios were evaluated: (i) no booster vaccination among staff or residents (baseline); (ii) booster vaccination administered to all residents but not staff; and (iii) booster vaccination administered to both staff and residents. Vaccination status was assigned at model initialisation and fixed throughout the 24-hour simulation cycle. All infection probability adjustments were applied based on vaccination status as previously described.

\section{Results}\label{sec:results}
\subsection{Sensitivity of Contact Patterns to Operational Parameters}\label{subsec:sensitivity}
Variation of operational parameters produced measurable differences in emergent contact structures. Across all scenarios, changes to simulation start time had negligible effects on contact volumes. By contrast, reducing the resident population from 60 to 45 led to lower total contacts, including a 12.4\% reduction in resident–resident contacts and a 9.8\% reduction in resident–staff contacts. Increasing the number of residents to 75 yielded proportionate increases in contact rates. The most substantial shifts were observed under changes in staff allocation. When staffing levels were increased to 3 staff per ward during day shifts (with 2 staff at night), staff-staff contact rates rose by 19.1\%, while staff-resident contact rates increased by 11.6\% compared to the base configuration. These results demonstrate that contact intensity within the facility is most sensitive to staff coverage levels, with comparatively limited influence from temporal scheduling or resident count.

\begin{table}[ht]
\centering
\caption{Sensitivity of average daily contact volumes to variation in start time, resident population, and staff-per-ward ratios. Each scenario was simulated 100 times under the specified operational parameters. Values represent the mean number of contacts per day by contact type: resident–resident, resident–staff, and staff–staff.}
\label{tab6}
\small
\begin{tabular}{p{3cm} c c c c c c}
\toprule
\textbf{Scenario} & \textbf{Start Time} & \textbf{Residents} & \textbf{Staff per Ward (1/2/3)} & \textbf{Res–Res} & \textbf{Res–Staff} & \textbf{Staff–Staff} \\
\midrule
Base & 6:15 & 60 & 2/2/1 & 60.2 & 128.9 & 109.5 \\
Start Time: 6:00 & 6:00 & 60 & 2/2/1 & 58.6 & 125.3 & 106.2 \\
Start Time: 6:30 & 6:30 & 60 & 2/2/1 & 62.1 & 130.7 & 112.8 \\
Residents: 45 & 6:15 & 45 & 2/2/1 & 56.6 & 122.5 & 102.5 \\
Residents: 75 & 6:15 & 75 & 2/2/1 & 65.3 & 135.2 & 116.7 \\
Staff per ward: 1 & 6:15 & 60 & 1/1/1 & 59.5 & 121.7 & 103.9 \\
Staff per ward: 3 & 6:15 & 60 & 3/3/2 & 66.1 & 139.4 & 119.6 \\
\bottomrule
\end{tabular}
\end{table}

\subsection{Contact Matrix}\label{subsec:contactmatrixresults}
\begin{figure}[H]
    \centering
    \includegraphics[width=0.75\linewidth]{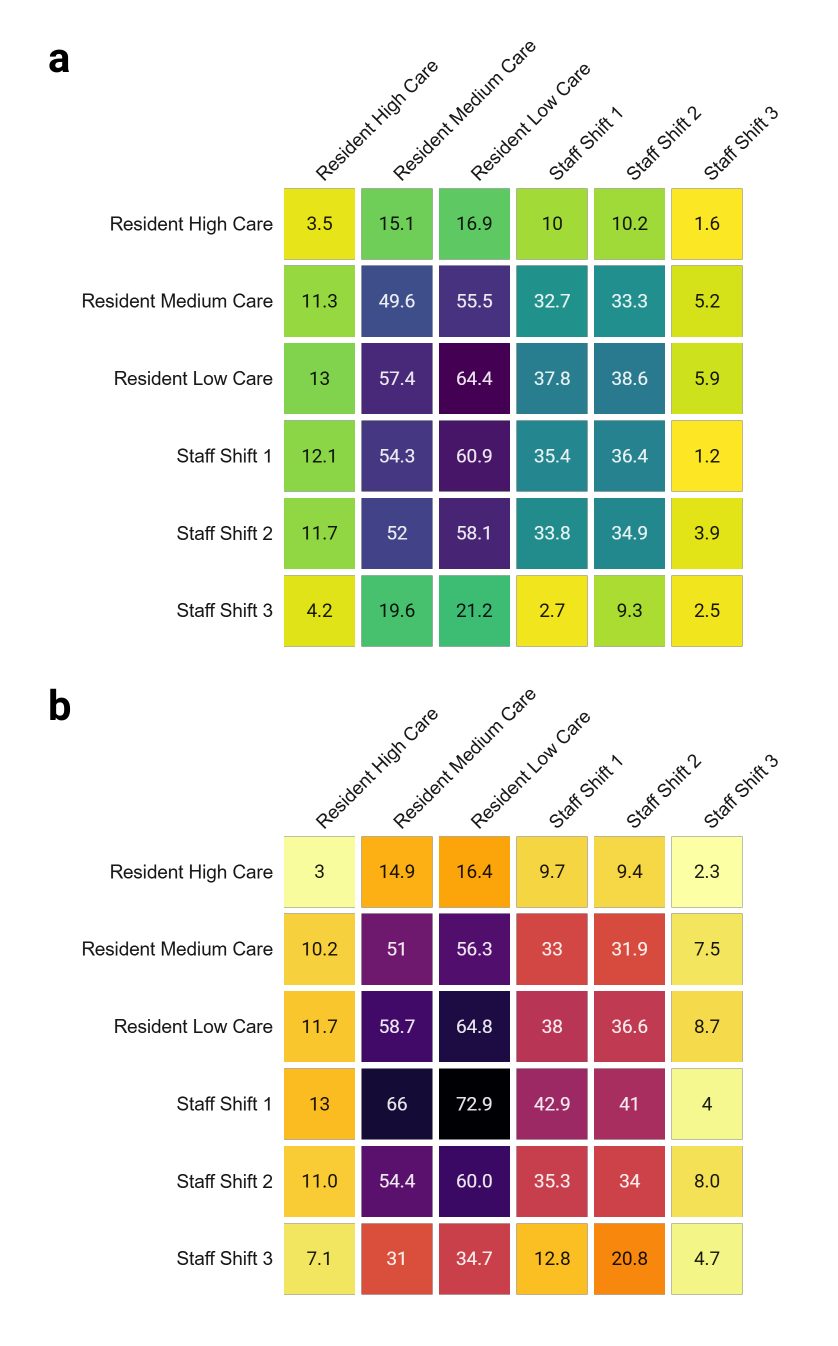}
    \caption{Predicted contact matrix showing mean contacts per person per day between agent type categories: high care residents, medium care residents, low care residents, and staff on shifts 1–3, for contact radii of 1.5 metres (A) and 3 metres (B). Contact rates were estimated using a generalised linear model with a Poisson distribution to adjust for differences in group size and stochastic variability.
}
    \label{fig4}
\end{figure}

For the contact matrix generation, the simulation ran for 5 days, starting at 6:15 AM on day 1 and ending at 6:15 AM on day 6. During this period, multiple data series were collected for the analysis, specifically focusing on two different contact radii: 1.5 meters and 3 meters [Figure \ref{fig4}].
At a 1.5-meter contact radius, simulated interactions among residents showed variation across different care levels. Residents in low care had the highest predicted intra-group interactions, averaging 64.4 contacts [95\% UI: 52.5–80.1], followed closely by medium care residents with 49.6 contacts [95\% UI: 41.7–56.3]. High care residents had substantially fewer intra-group contacts, averaging 3.5 [95\% UI: 2.3–5.0]. Staff-resident interactions were most frequent for shift 1 and shift 2 care workers, particularly with medium care residents (60.9 [95\% UI: 45.0–73.3] and 58.1 [95\% UI: 33.3–72.0] contacts, respectively). Contacts with low care residents were also high for shift 1 (54.3 [95\% UI: 41.3–61.9]) and shift 2 (52.0 [95\% UI: 28.2–64.6]). In contrast, shift 3 staff demonstrated the lowest levels of interaction across all resident types, with resident contacts not exceeding 35 contacts across any group and intra-shift contacts averaging only 2.5 [95\% UI: 0.6–6.4].

At a 3-meter contact radius, the simulated interaction patterns largely mirrored those observed at 1.5 meters, though overall contact frequencies were slightly increased in most groups. Low care residents had the highest intra-group interactions, averaging 64.8 contacts [95\% UI: 45.3–69.1], followed closely by medium care residents with 51.0 contacts [95\% UI: 47.8–52.1]. High care residents continued to exhibit fewer intra-group contacts, averaging 3.0 [95\% UI: 2.5–4.6]. Among staff, shift 1 had the most frequent predicted interactions, particularly with medium care residents (72.9 [95\% UI: 49.3–81.2]) and low care residents (66.1 [95\% UI: 44.8–72.1]). Shift 2 staff showed similar patterns, with 60.0 contacts with medium care residents [95\% UI: 49.9–65.2] and 54.4 with low care residents [95\% UI: 52.5–59.1]. Shift 3 staff maintained the lowest levels of interaction across all groups, with resident contacts not exceeding 36 contacts and intra-shift contacts averaging only 4.7 [95\% UI: 4.2–6.0].

For the 1.5-meter contact radius, the GLM analysis applied to the simulated contact data revealed that low care residents had a significantly greater likelihood of being involved in interactions compared to high care residents, with a coefficient of 1.35 [95\% CI: 1.07 to 1.58. Medium care residents also showed elevated predicted interaction frequencies, with a coefficient of 1.21 [95\% CI: 0.97 to 1.37]. Among staff, shift 1 and shift 2 personnel were predicted to have higher interaction likelihoods, with coefficients of 1.32 [95\% CI: 0.79 to 1.60] and 1.26 [95\% CI: 0.58 to 1.64], respectively. Staff on shift 3 had the lowest predicted interaction frequency among staff categories, with a coefficient of 0.74 [95\% CI: 0.42 to 1.07]. On the contacting side, residents in low care and medium care were strong predictors of increased contact counts, with coefficients of 1.52 [95\% CI: 1.19 to 1.75] and 1.62 [95\% CI: 1.28 to 1.91], respectively. Contact initiated by shift 3 staff was associated with a reduced frequency of interactions, reflected in a negative coefficient of –0.38 [95\% CI: –0.76 to 0.02].

For the 3-meter contact radius, the GLM analysis averaged across all simulation runs revealed increased interaction likelihoods for low and medium care residents compared to the high care reference group. Specifically, low care residents had a mean coefficient of 1.40 [95\% UI: 1.09 to 1.52] , while medium care residents showed a higher interaction likelihood with a coefficient of 1.25 [95\% UI: 0.99 to 1.36]. Staff on shift 1 exhibited the highest predicted interaction frequency among contactees, with a coefficient of 1.52 [95\% UI: 1.06 to 1.69], followed by shift 2 staff with 1.31 [95\% UI: 1.10 to 1.48]. Shift 3 staff showed no clear effect, with a coefficient near zero (0.01 [95\% UI: -0.60 to 0.94]). When examining interaction frequency from the perspective of the contacter, low care and medium care residents again showed elevated coefficients of 1.65 [95\% UI: 1.29 to 1.82] and 1.76 [95\% UI: 1.36 to 1.94], respectively. Staff shift 1 and shift 2 also showed moderate increases in contact likelihood as originators of contact (1.22 [95\% UI: 0.84 to 1.40] and 1.17 [95\% UI: 0.91 to 1.30], respectively). In contrast, shift 3 staff had a negative coefficient of -0.84 [95\% UI: -1.35 to -0.13], indicating a substantially lower propensity to initiate contact events.

\subsection{Staff and Resident Contact Analyses}\label{subsec:staffandresidentanalysis}
\begin{figure}[H]
    \centering
    \includegraphics[width=1\linewidth]{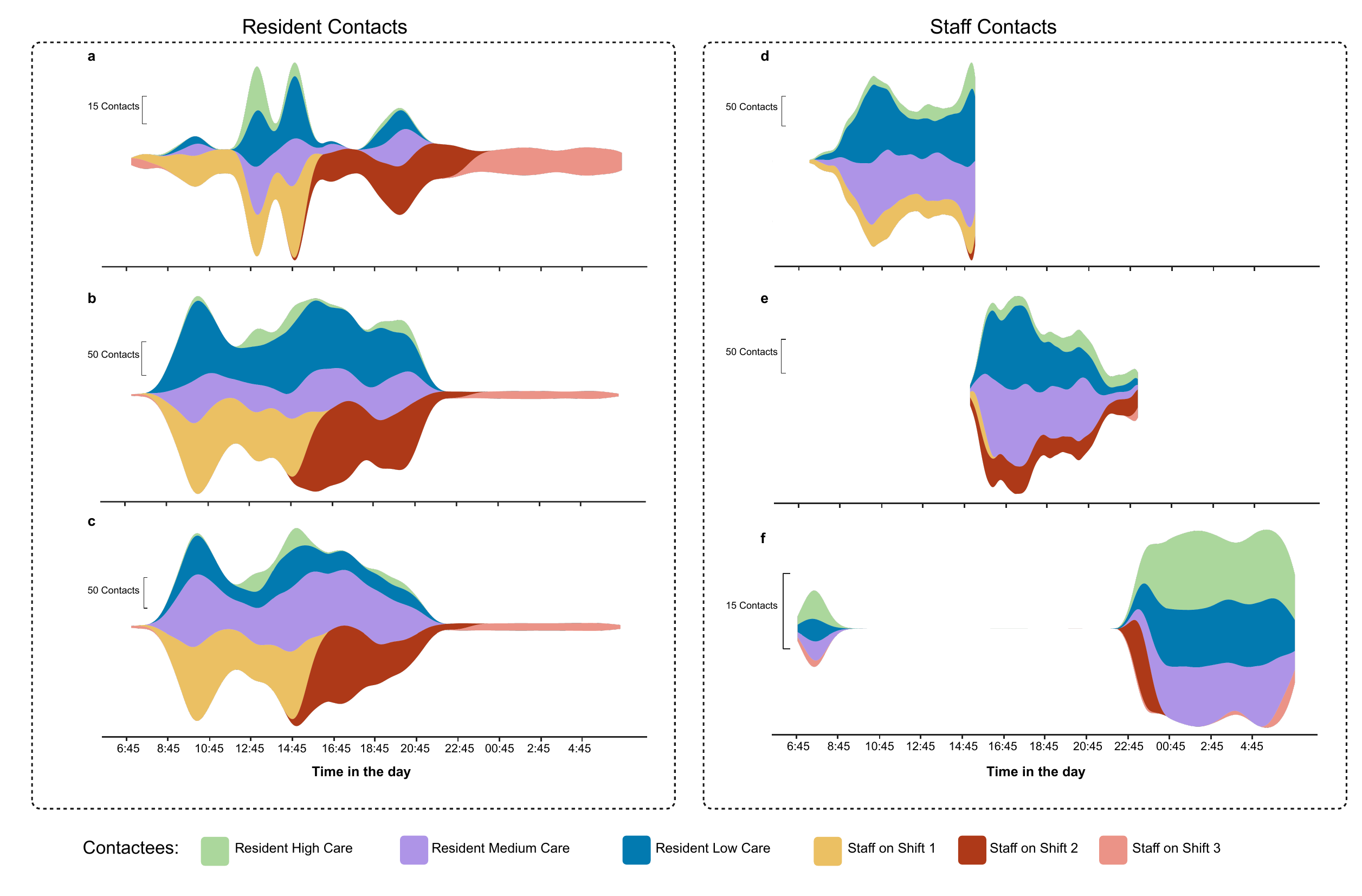}
    \caption{Rolling 20-minute contact totals for residents and staff across the day. The plots show how the number of contacts changes over time for high care (a), medium care (b), and low care residents (c), and for staff on shifts 1 (d), 2 (e), and 3 (f). Each point reflects the total number of contacts recorded in the previous 20 minutes, updated at one-minute steps. Colours indicate which agent types the contacts involved, as shown in the legend.}
    \label{fig5}
\end{figure}

To examine temporal variation in interaction intensity, we calculated a 20-minute rolling sum of contacts for each agent category. The rolling sum represents the total number of contacts involving that agent category within the preceding 20-minute window, updated at each one-minute timestep to produce a smoothed time-series of activity [Figure \ref{fig5}]. For residents, there is variability in both peaks of frequency and sustained contacts between care levels. High care residents demonstrated lower overall predicted contact frequency, with a mean rolling sum of 0.87 contacts [95\% CI: 0.82 to 0.92]. The maximum recorded contact for this group was 22. Medium care residents exhibited substantial predicted contact activity throughout the day, with a mean rolling sum of 3.99 [95\% CI: 3.82 to 4.16]. The maximum contact peak for this group was 66. Low care residents also showed significant predicted contact activity, with a mean rolling sum of 3.80 [95\% CI: 3.63 and 3.96], with a maximum contact peak of 65.
The staff contact patterns varied across the three shifts with shift 1 care staff experiencing a predicted mean rolling sum of 7.62 contacts [95\% CI: 7.27 to 7.97], with a maximum contact recorded of 47. Shift 2 staff had a predicted mean rolling sum of 5.94 [95\% CI: 5.64 to 6.23], with a maximum of 46. Staff in Shift 3 demonstrated the lowest predicted contact frequency, with a mean rolling sum of 0.52 [95\% CI: 0.49 to 0.54], and a maximum of 15.

\begin{figure}[H]
    \centering
    \includegraphics[width=0.8\linewidth]{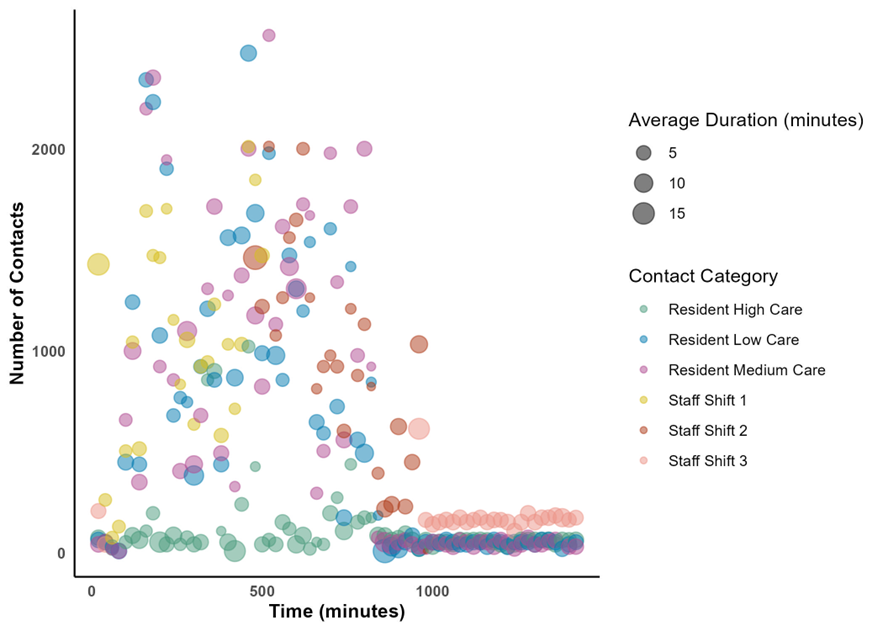}
    \caption{Predicted contacts by time across a 24-hour simulation cycle, summarised in 20-minute intervals for residents (by care level) and staff (by shift). Each bubble represents the number of contacts in that interval; bubble size indicates the average contact duration in minutes.}
    \label{fig6}
\end{figure}

Predicted contact volumes varied over the course of the simulation day and differed across contactee categories. Contact counts exhibited clear temporal structure when aggregated into 20-minute intervals and averaged across stochastic simulation runs. Periods of increased contact frequency coincided with specific times of day, with variation in both the number and duration of contact events across categories [Figure~\ref{fig6}].

Predicted contacts involving high care residents were relatively stable and evenly distributed throughout the simulation period, with durations ranging from short to extended interactions. In contrast, contacts involving medium and low care residents exhibited greater temporal variability, including pronounced peaks at specific times that may reflect structured daily routines or care demands.

Staff contact patterns were temporally distinct across shifts. Staff in shifts 1 and 2 maintained a broad and consistent distribution of predicted contacts, whereas shift 3 demonstrated a marked increase in contact frequency later in the simulation period, potentially corresponding to end-of-day tasks or workload redistribution. Notably, shift 3 also showed elevated average durations during interactions with medium care residents.

Clusters of longer-duration predicted contacts were observed primarily among high care residents and staff in shift 3, suggesting intervals of intensified interaction. Additionally, sporadic outliers with unusually high contact frequencies or durations were evident across all categories, indicating rare but potentially important deviations from typical patterns that may warrant further examination.

\begin{figure}[H]
    \centering
    \includegraphics[width=1\linewidth]{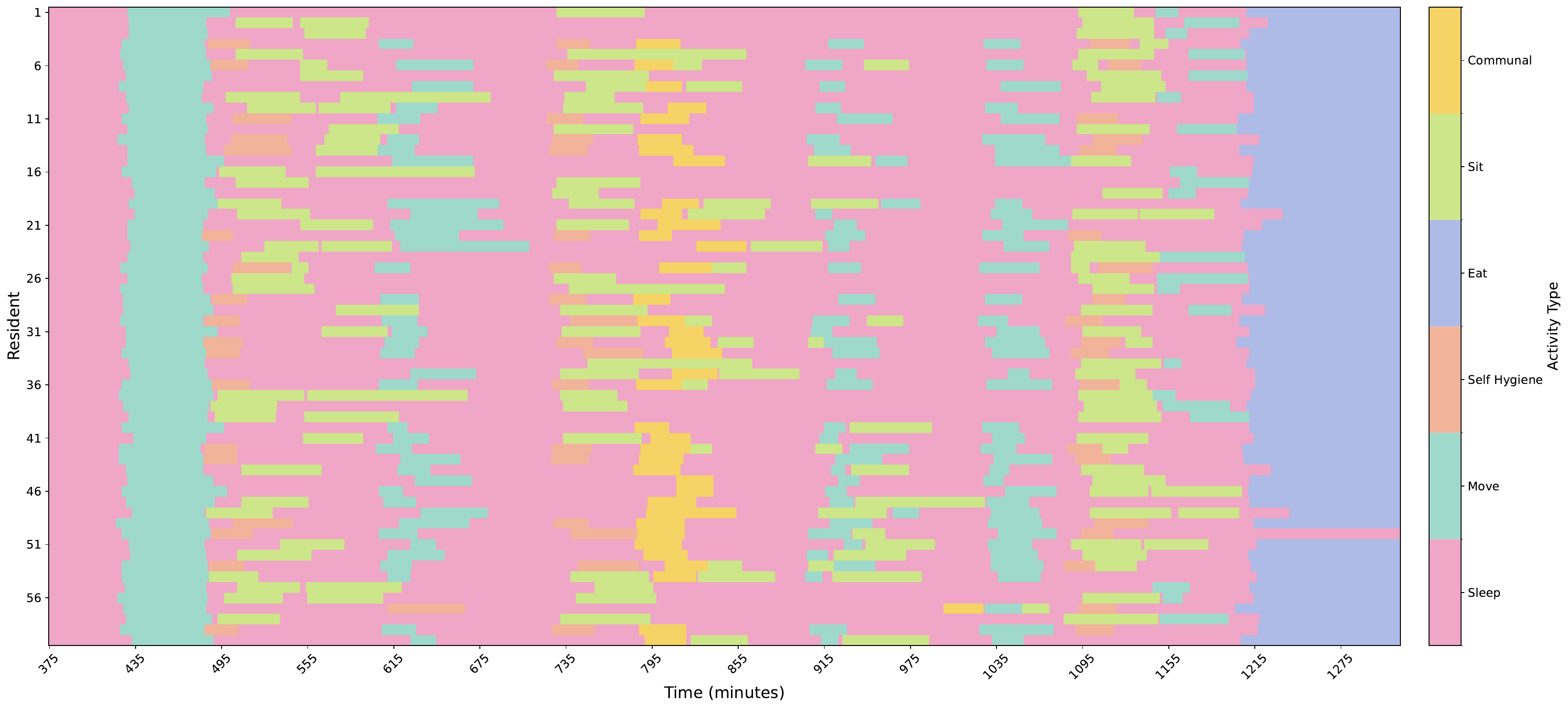}
    \caption{Predicted representative patterns of residents over a 24-hour period, illustrating the distribution of time spent on various activities.}
    \label{fig7}
\end{figure}

Across the simulated day, residents spent the majority of time in sitting or sleep states, with other activities occurring in shorter, structured intervals. Communal activities were concentrated within distinct time blocks, whereas self-hygiene events were infrequent and confined to narrow windows. Sleep activity was predominantly clustered overnight, with inter-individual variation in timing and duration. Periods of movement between destinations were brief and distributed throughout the day, typically occurring around transitions between activities. Mealtime activity was temporally dispersed across residents rather than occurring at a single synchronised time [Figure \ref{fig7}].

\subsection{Transmission Simulation}\label{subsec:transmissionsim}
\begin{figure}[H]
    \centering
    \includegraphics[width=0.75\linewidth]{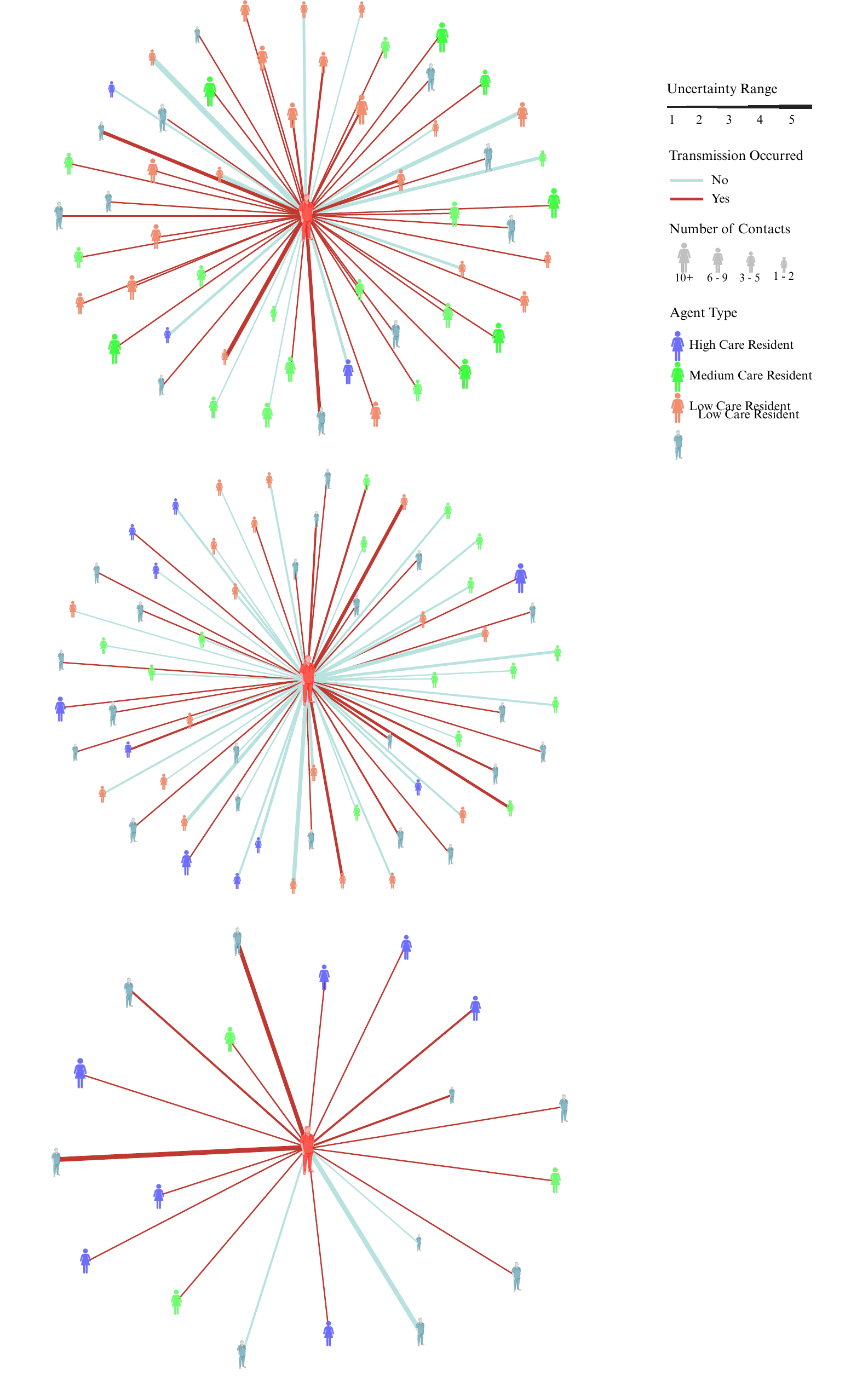}
    \caption{Transmission networks for each staff shift following seeding of a single infectious staff member. Nodes represent individual agents and are scaled by the total number of contacts made during the simulation. Edges represent aggregated contact events between the infectious staff member and other agents, coloured red when transmission occurred and blue otherwise. Edge length corresponds to cumulative contact duration, while edge width reflects uncertainty in the estimated transmission probability (95\% CI).}
    \label{fig8}
\end{figure}

To assess baseline transmission risk within the facility prior to vaccination interventions, infection events were simulated independently for each staff shift by seeding a single infectious staff agent per 24-hour simulation cycle. For each shift, 100 stochastic simulations were conducted with the same seeding conditions, and results represent the mean number of transmission events across these runs. Transmission probabilities were dynamically evaluated based on recorded contact durations, with airborne transmission uncertainty incorporated through 95\% confidence intervals reflecting variation in quanta generation and inhalation rates. Across all shifts, transmission risk increased with cumulative exposure time, with longer-duration contacts associated with higher median and upper-bound transmission probabilities [Figure \ref{fig8}].

A total of 304 contacts were recorded during Shift 1, resulting in 73 transmission events and a transmission occurrence of 24.0\% [95\% CI: 21.0\% - 28.5\%]. The mean contact duration across all shift 1 contacts was 2.6 minutes, while contacts resulting in transmission had a substantially higher average duration of 8.8 minutes compared to 0.6 minutes for non-transmitted contacts. Transmission events predominantly involved interactions between the infectious staff member and medium care residents, accounting for 45.2\% of all recorded infections.

During Shift 2, 247 contacts were recorded, with 45 transmission events and a transmission occurrence of 18.2\% [95\% CI: 15.2\% - 22.0\%]. The average contact duration for all contacts was 2.5 minutes. Transmitted contacts had an average duration of 12.0 minutes, whereas non-transmitted contacts averaged 0.4 minutes. Transmission was more evenly distributed across resident care levels compared to shift 1, although medium care residents remained the most frequent recipients of infection.

Shift 3 recorded a total of 122 contacts and 52 transmission events, corresponding to a transmission occurrence of 42.6\% [95\% CI: 38.0 - 47.5\%], the highest proportion observed across shifts. The mean duration of contacts was 4.3 minutes overall, with transmitted contacts averaging 7.3 minutes and non-transmitted contacts averaging 2.0 minutes. High care residents constituted the largest proportion of infections at 52.9

Wider uncertainty bounds were observed for transmission probabilities associated with single contacts of extended duration, particularly during shift 1. For contacts exceeding 30 minutes of cumulative exposure, upper-bound transmission probabilities were consistently elevated, with greater spread between the lower and upper 95\% confidence intervals compared to shorter contacts.

\begin{figure}[H]
    \centering
    \includegraphics[width=0.8\linewidth]{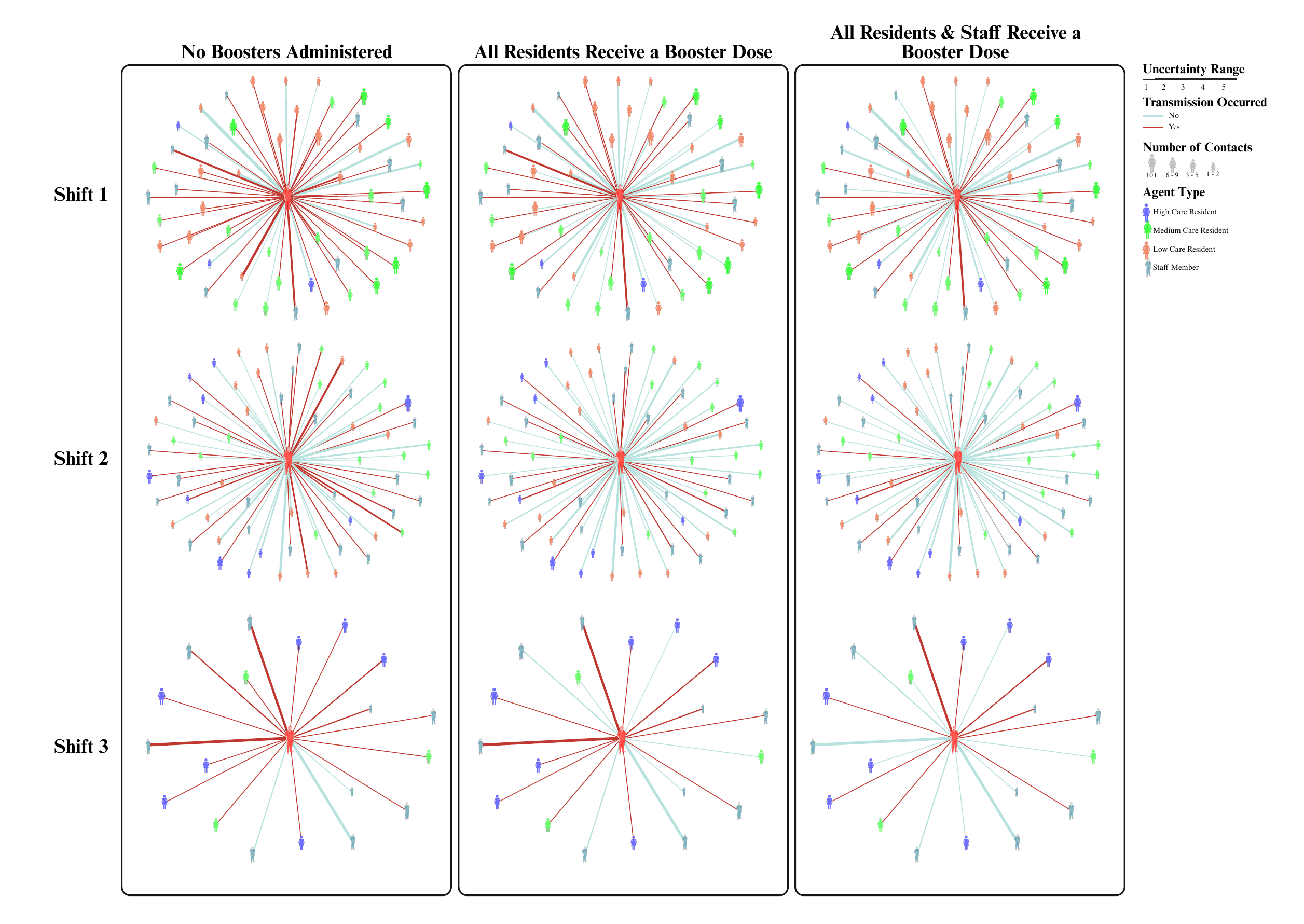}
    \caption{Network representation of contact and transmission outcomes during each staff shift under vaccination intervention scenarios following seeding of a single infectious staff member. Nodes correspond to individual agents and are scaled by the total number of contact events involving that agent across the simulation period. Edges represent the aggregated contact history between the infectious staff member and each contacted agent, with edge colour indicating whether transmission occurred (red) or did not occur (blue). Edge length is proportional to the cumulative duration of contact between the pair, while edge width reflects the uncertainty (95\% confidence interval) associated with the estimated transmission probability after adjustment for vaccination status. Separate networks are shown for Shift~1, Shift~2, and Shift~3.}
    \label{fig9}
\end{figure}

To assess the impact of booster vaccination on transmission risk, two intervention scenarios were simulated and compared against the baseline model, representing waned immunity with no effective protection. Scenario 1 implemented booster vaccination among all residents only, while Scenario 2 implemented booster vaccination among both residents and staff. Across all shifts, the introduction of vaccination reduced the number of transmission events compared to the baseline scenario, with greater reductions observed when both residents and staff were vaccinated [Figure \ref{fig9}].

For Shift 1, the baseline simulation recorded 73 transmission events out of 304 total contacts. Under Scenario 1, where only residents were vaccinated, transmission events decreased to 34, representing a 53.4\% reduction. Scenario 2, with both residents and staff vaccinated, further reduced transmission events to 25, corresponding to a 65.8\% reduction compared to baseline.

In Shift 2, 45 transmission events were recorded under the baseline model from 247 total contacts. Vaccinating residents alone (Scenario 1) reduced transmission events to 35, representing a 22.2\% reduction. The addition of staff vaccination (Scenario 2) led to a further decrease, with only 14 transmission events recorded, which is a 68.9\% reduction relative to baseline.

For Shift 3, where baseline conditions resulted in 52 transmission events out of 122 total contacts, Scenario 1 reduced transmission to 23 events (a 55.8\% reduction). Scenario 2 achieved a greater reduction, with 17 events recorded, corresponding to a 67.3\% decrease compared to the baseline.

\section{Discussion}\label{sec:discussion}
This study demonstrates the value of ABMs in capturing the complex interaction patterns that traditional survey-based methods might overlook, particularly in specialised environments like aged care homes \cite{mossong2008social,wallinga2006socialcontacts,hoang2019contactsurveys}. 

The results from the contact patterns at both 1.5-meter and 3-meter radii emphasise the significant variability in interactions based on the care level of residents. The model predicted that low care residents would have the highest frequency of contacts, both among themselves and with staff. This finding is particularly noteworthy as it suggests that low care residents, who retain greater mobility and are more socially active, function as critical nodes in the contact network within aged care facilities. Given their centrality in the contact matrix, these residents could potentially play a significant role in the transmission of infectious diseases within the facility. Therefore, infection control measures might need to be particularly stringent for this subgroup, focusing on minimising unnecessary contacts and implementing targeted interventions during high-contact activities. 

In contrast, the simulation predicted that high care residents, who have limited mobility, would have significantly fewer interactions, primarily involving contact with staff rather than with other residents. This observation is consistent with the expected behaviour in aged care settings, where residents with higher care needs are often more isolated and dependent on staff for assistance. The reduced contact among high care residents indicates a lower likelihood of resident-to-resident transmission within this group, shifting the focus of infection control to staff-resident interactions, particularly during personal care activities. This dependency highlights the importance of ensuring that staff are adequately trained in infection prevention protocols and that there is strict adherence to hygiene practices during these interactions \cite{mossong2008social,wallinga2006socialcontacts}.

The variation in interaction patterns across different staff shifts further illustrates the complex dynamics at play within aged care facilities. The first two shifts were associated with the highest contact frequencies, corresponding with peak periods of resident activity and staff engagement. This suggests that these shifts are critical windows for potential disease transmission, particularly among more active residents and staff. In contrast, the third shift exhibited markedly lower interaction rates, likely reflecting the reduced staff-to-resident ratio and the overall decrease in resident activity during nighttime hours. These findings highlight the necessity for infection control measures to be dynamically adjusted according to the time of day, with heightened vigilance and resources allocated during periods of increased activity.

The GLM analyses further reinforces these observations by quantitatively assessing the likelihood of contacts across different resident and staff categories. The significant coefficients associated with low care residents and the first two staff shifts highlight the increased propensity for contact initiation within these groups, suggesting they are pivotal in the network of interactions within the facility. This insight is crucial for the design of infection prevention strategies, which should prioritise these high-contact groups to effectively curb the spread of pathogens. Conversely, the non-significant coefficient for shift 3 staff may indicate a lower impact on overall contact dynamics, potentially due to the nature of tasks performed during this shift, which may involve less direct interaction with residents.	

Furthermore, the model's flexibility extends to the simulation of communal spaces and their use by different resident groups at varying times. By capturing these dynamics, the model provides a more comprehensive understanding of how spatial and temporal factors influence contact patterns and potential disease transmission within a facility. As the behavioural logic is decoupled from the specific floorplan, the framework can be applied to facilities with alternative spatial configurations, allowing these spatial dynamics to be explored in a range of architectural contexts. This feature allows for the exploration of various "what-if" scenarios, such as the effects of altering staff shift patterns or reorganising communal spaces to minimise high-risk interactions. The insights gained from such simulations can inform the implementation of more effective infection control measures, tailored to the specific needs and configurations of each facility. Future research could build on this work by integrating the contact matrix generated, in this study, with a deterministic multicompartment epidemic model. This integration would enable a deeper exploration of how specific changes in contact patterns, influenced by shifts in staff schedules or the reorganisation of communal spaces, impact the transmission dynamics of infectious diseases. Such a combined approach could provide more precise estimates of the effectiveness of targeted interventions, offering a powerful tool for optimising infection control strategies in aged care settings. By linking the detailed interaction data from the ABM with the predictive capabilities of a deterministic model, researchers could advance our understanding of the interplay between behavioural dynamics and disease spread, paving the way for more informed public health decisions in these critical environments.

The integration of a probabilistic transmission model refined the interpretation of contact patterns by linking cumulative interaction duration to estimated infection risk. Transmission probabilities rose with longer exposures, consistent with the exponential structure of the modified Wells-Riley equation. The frequency and temporal depth of contacts both contributed meaningfully to infection potential. Shift 1, which involved sustained and high-density interactions, produced the greatest number of predicted infections. In contrast, shift 3 exhibited markedly lower risk, reflecting reduced contact intensity and shorter exposure durations, which could be attributed to staff numbers and night shift schedule. Incorporating uncertainty in airborne infectiousness and inhalation rates revealed that longer interactions also produced wider confidence intervals, highlighting the sensitivity of high-risk encounters to transmission parameter assumptions.

To assess mitigation, two booster vaccination scenarios were evaluated: resident-only vaccination and combined resident and staff vaccination. In both scenarios, vaccination reduced the probability of infection per contact, scaled by point estimates of vaccine effectiveness with associated 95\% confidence intervals. Across all shifts, transmission events declined following vaccination, with larger reductions observed when both residents and staff were immunised. Shifts 1 and 3, which had higher baseline transmission levels, showed the greatest impact. In shift 2, where contact rates were lower, the effect of resident-only vaccination was modest, though a further decline was observed with full vaccination coverage. As seeding and contact conditions were held constant across scenarios, observed reductions could be attributed directly to differences in vaccination status. Including parameter uncertainty enabled clearer interpretation of shift-specific variability in vaccine impact.

There are several limitations to this study that warrant consideration. First is the absence of empirical validation against direct observation or sensor-based measurement of contact patterns, such as wearable RFID technology. While these technologies can offer fine-grained insight into interpersonal proximity, they present substantial challenges in aged care settings, including cost, ethical concerns, and feasibility of deployment among residents with cognitive or mobility impairments. Consequently, no such dataset currently exists for this environment. Instead, this study relied on a survey of aged care workers to capture daily activity patterns, task timings, and resident interaction frequencies. Although this introduces reliance on self-reporting and inferred contact proxies, it represents a more context-specific and grounded approach than using generic population-level matrices. In addition, the findings presented reflect a single run of the agent-based model under a fixed set of conditions. Given the inherent stochasticity in agent-based models, repeated experimentation using different pseudorandom seeds would allow a more comprehensive understanding of variability in outcomes and support assessment of the robustness of emergent patterns. A further limitation is that the transmission component represents only primary infections within a single 24-hour period, without modelling secondary or subsequent transmission, meaning that potential multi-day outbreak trajectories are not captured in the present analysis. Finally, although Emergency Response behaviours are included in the task structure to represent acute clinical events or facility-wide disruptions, no emergency triggers were scheduled in the scenarios used here.

Future extensions of the model could incorporate scenarios with varying numbers of infectious staff across shifts, together with different staff and resident vaccination combinations. Expanding the simulation design in this way would allow systematic exploration of interactions between seeding intensity, shift structure, and immunity status, and would enable the use of regression-based analyses to quantify the relative contribution of each factor to overall transmission risk.

\section{Conclusion}
In conclusion, ABMs allow for more precise predictions of resident-staff interactions, enabling the design of highly targeted infection control strategies in a safe, adaptable, reproducible and detailed synthetic environment. The ability to tailor these models to the specific conditions of each facility, whether through variations in staff shifts, resident activity patterns, or architectural layouts, ensures that they can effectively address the unique challenges posed by different settings. As aged care facilities continue to face evolving threats from infectious diseases, the integration of such sophisticated modelling techniques is essential for safeguarding the health of both residents and staff.

\section{Ethics Statement}
Ethics approval from The University of New South Wales Human Research Ethics Committee (approval number: HC230501) and the aged care provider’s Research and Governance Office approval (Project code: R277). Survey design, data categorisation and details on informing parameters can be found within Appendix A.

\section{Code Availability}
The full AnyLogic model, including the facility layout module and behavioural logic, is openly available at: \\
\url{https://github.com/BREATHE-UNSW/AgedCareContactMatrix}

\section{Author Contribution Statement}
Conceptualisation and study aims were developed by HS, DH and CRM. Survey design was performed HS, CP and MK under the supervision of DH and CRM. Ethics and the aged care research governance applications for the survey were performed by HS under the supervision of DH, CP and CRM. Model development, parameterisation, validation and calibration was performed by HS under the supervision of DH. Contact matrix generation and GLM analyses were performed by HS under the supervision of DH. Finally, this study was written by HS and reviewed by DH, MK, CP and CRM.

\bibliographystyle{plainnat}
\bibliography{references}
\section{Appendix A: Survey}
Appendix A provides detailed survey instruments and parameterisation rationale supporting the main Methods section.
\subsection{Study Design}
A cross-sectional survey was conducted in 2023 to parameterise an agent-based model (ABM) aimed at creating a contact matrix for use in infectious disease transmission models within aged care settings in Sydney, Australia. The primary objective was to address the absence of standardised contact matrices for aged care environments, which are essential for accurately modelling disease transmission and enhancing outbreak preparedness. The survey focused on collecting comprehensive data on the interactions between staff and residents within a single aged care provider, with the goal of developing a robust contact matrix to be utilised in agent-based models.
The survey targeted aged care workers across various shifts within the facilities, gathering detailed information on their daily interactions with residents. Participants were asked to complete a structured 21-question survey designed to capture demographic data, care environment details, resident characteristics, contact and interaction patterns, and activity and mobility data. The survey's design was informed by established methodologies that emphasise the importance of capturing both temporal aspects of daily routines and broader contextual factors in care work \cite{crosbie2006using,nilsson2009registered, sveinsdottir2018content}.
An activity diary component was included, allowing participants to log their activities and interactions in real-time, which was effective in documenting the repetitive and routine nature of care \cite{crosbie2006using}. This provided a precise account of their daily routines. Structured questions complemented the diaries by gathering demographic data, care environment details, and resident characteristics, ensuring a comprehensive understanding of the factors influencing care practices \cite{nilsson2009registered}.
The diary entries were organised to capture relevant details about work activities, including the timing, duration, and nature of interactions with residents, aligning with recommendations to ensure thorough data collection \cite{sveinsdottir2018content}. All responses were anonymised and securely stored, and the data were subsequently used to parameterise the agent-based model (ABM).
Ethical approval for the study was obtained from the provider’s Research and Governance Office (Project code: R277) and The University of New South Wales Human Research Ethics Committee (approval number: HC230501).

\subsection{Recruitment and Participants}
Recruitment for the survey was conducted within aged care provider facilities through the distribution of recruitment flyers displayed in staff rooms. These flyers included a QR code that directed staff to a detailed explanation of the study and provided access to the electronic survey. To accommodate the diverse linguistic backgrounds of the staff, the survey was made available in English, Nepalese, and Filipino.
Staff members willing to participate were invited to complete the survey electronically during their off time. The recruitment flyer also provided a link to the Participant Information and Consent Form (PICF), which participants could access and sign electronically before completing the survey.

\subsection{Data Parameterisation}
Each survey question was specifically designed to gather information on the tasks and interactions occurring during a staff member’s shift, with a focus on the timing, frequency, and duration of these activities and the broader environmental settings.
This data allowed for the assignment of specific times and durations to tasks, such as aiding in self-care, meal preparation, and medication runs. These tasks are critical for accurately modelling the daily routines and interactions within the aged care environment. By incorporating this detailed data into the agent-based models, the study aimed to create a contact matrix that accurately mirrors the conditions within the facility. This precise representation is vital for predicting the spread of infectious diseases and improving outbreak preparedness, ensuring that the model closely reflects the dynamics of an aged care setting.

\subsection{Survey Responses and Parameterisation}
\subsubsection{Shift Schedules and Staffing Levels}
The survey provided data on the typical shifts worked by staff members, which directly informed the model's shift schedules. The model was parameterised to reflect the start and end times of each shift, ensuring alignment with the patterns reported by the staff. Shift 1, which had the highest staffing level (57\% of participants), was particularly influential in shaping the model’s workload distribution. Shift 2 and Shift 3, with fewer staff members, were also integrated into the model to simulate varying levels of activity and care delivery throughout the day. 

\subsection{Resident Care Levels and Staff Assignments}
The survey also informed how residents' care levels are managed by the staff, which was used to assign staff to specific care levels within the model. The model was parameterised to reflect the distribution of care responsibilities, with staff being assigned to high care, medium care, or low care residents based on the survey responses. For example, 52\% of staff worked with a combination of high care and medium care residents, while others worked with medium and low care residents. Task prioritisation, task duration, and staff allocation varied according to the care level of the resident. 

\subsubsection{Task Durations and Frequencies}
Survey responses regarding the time spent on various tasks, such as meal preparation, medication runs, and aiding in self-care, were used to inform the model’s task durations and timing. The model was parameterised to reflect the average time spent on these tasks, with confidence intervals incorporated to introduce variability. For example, meal preparation for lunch and dinner was modelled with an average duration of 35.83 minutes, based on the conditions reported by staff. Breaks and downtime were also carefully integrated into the model based on survey data, reflecting the typical break times reported by staff between the 6th and 7.5th hour of their shift. This ensured that the model mirrored the staffing strategies reported by the staff, where only one staff member from each area is on break at any given time.
The frequency of communal activities, as reported in the survey, informed the scheduling of these activities within the model. The model was parameterised to simulate how often and when communal activities take place, ensuring that staff were appropriately allocated to assist residents during these times. This parameterisation was essential for accurately reflecting the social interactions and care routines that occur within the facility.
\subsubsection{Task Prioritisation and Allocation}
Task prioritisation, duration, and staff allocation varied according to the care level of the resident, as revealed by the survey responses. Open-ended questions such as "Can you describe your resident's morning routine, including waking up, personal care activities, and self-care? How long do you normally spend with each resident's self-care?" provided detailed insights into how time is allocated, where interactions take place, and which tasks are prioritised.
For example, residents requiring higher levels of care often necessitate more time and assistance, especially during personal hygiene activities. The responses indicated that staff often spend significant time with these residents, sometimes requiring two staff members to assist, particularly in cases involving physical or cognitive challenges. These responses highlighted the need to allocate more staff and time to tasks involving high-care residents.
The survey responses also shed light on the environmental context of these interactions. For instance, residents with limited mobility were primarily assisted in their rooms or bathrooms, whereas those with higher mobility might engage in personal care activities in communal areas. This variation in the interaction environment underscores the importance of considering the specific locations where care is provided when allocating tasks and staff.
Furthermore, structured questions regarding the activities performed during each hour of the shift provided concrete data on task durations and frequencies. For example, respondents detailed the time spent on tasks such as meal preparation, medication administration, and personal care, which can be used to estimate the average duration of these activities. Additionally, insights into staff break times and the scheduling of communal activities offer crucial data for planning staff coverage and ensuring that residents receive the necessary attention during these periods.
Finally, survey responses about the proximity and frequency of interactions between staff and residents, along with the number of residents requiring different levels of assistance during meals, help to define contact patterns. For example, staff often reported being in close proximity to residents during feeding times, with varying levels of required assistance, which can inform how close contacts are defined in terms of distance and duration.
\subsubsection{Contact Definitions and Interaction Patterns}
The survey provided key insights into how frequently staff interacted with residents and the proximity of these interactions, which were used to define the contact parameters within the model. Contacts were modelled based on the distance between individuals and the duration of the interactions, with significant contacts being defined as those lasting 3 seconds or more. This information was crucial for constructing the contact matrix within the model, which is used to simulate potential points of disease transmission within the facility.
The allocation of staff to specific tasks, particularly those likely to result in close contact with residents, was also informed by the survey data. The model was parameterised to reflect which tasks, such as aiding in self-care or mealtime assistance, were most likely to involve close contact, thereby influencing the opportunities for disease transmission.

\subsubsection{Resident Activities and Preferences}
The survey data on residents' mobility and preferred locations for activities provided essential information for parameterising resident behaviours within the model. The model was designed to simulate where residents chose to sit, eat, or participate in activities, ensuring that these behaviours aligned with the patterns reported by staff. For example, residents with lower mobility were modelled to prefer locations closer to their living areas, while more mobile residents had a broader range of locations.
Variation in task execution, such as the time taken to assist with self-care or prepare meals, was also informed by the range of durations reported in the survey. This allowed the model to introduce realistic variability in how tasks were performed, capturing the diverse ways in which care is delivered within the facility.

\subsubsection{Environmental Setup and Population Distribution}

The staff-to-resident ratios, as reported in the survey, were used to parameterise the model’s staffing levels within each care level. This ensured that the simulation reflected the workload and staff distribution, providing a realistic foundation for the model's operations.
The environmental layout and functionality within the model were also informed by the survey, particularly regarding where tasks are typically performed. The model was designed to replicate the spatial organisation of the facility, including the specific wings or areas where staff and residents are likely to interact. This spatial parameterisation was critical for accurately modelling the flow of activities and interactions within the aged care environment.
\end{document}